\documentstyle[aas2pp4]{article}

\def\simless{\mathbin{\lower 3pt\hbox
     {$\rlap{\raise 5pt\hbox{$\char'074$}}\mathchar"7218$}}}   
\def\simmore{\mathbin{\lower 3pt\hbox
     {$\rlap{\raise 5pt\hbox{$\char'076$}}\mathchar"7218$}}}   
\def\aql{Aquila~X~--~1}                                     

\received{}
\accepted{}
\slugcomment{Submitted to ApJ, December 2002}

\lefthead{Reig et al.}
\righthead{Timing properties and spectral states in Aquila~X~--~1}

\begin{document}

\title{Timing properties and spectral states in Aquila X--1}

\author{P.~Reig\altaffilmark{1},
S.~van~Straaten\altaffilmark{2},
M.~van~der~Klis\altaffilmark{2}
}

\altaffiltext{1}{G.A.C.E, Instituto de Ciencias de los Materiales, 
University of Valencia, 
P.O. Box 22085, E - 46071 Paterna-Valencia, Spain}

\altaffiltext{2}{Astronomical Institute ``Anton Pannekoek'',
       University of Amsterdam and Center for High-Energy Astrophysics,
       Kruislaan 403, NL-1098 SJ Amsterdam, the Netherlands}

\begin{abstract}

We have analyzed five X-ray outbursts of the neutron-star soft X-ray
transient \aql\ and investigated the timing properties of the source in
correlation with its spectral states as defined by different positions in
the color-color and hardness-intensity diagrams. The hard color and the
source count rate serve as the distinguishing parameters giving rise to
three spectral states: a low-intensity hard state, an intermediate state
and a high-intensity soft state. These states are respectively identified
with the extreme island, island and banana states that characterize the
atoll sources. The large amount of data analyzed allowed us to perform for
the first time a detailed timing analysis of the extreme island state.
Differences in the aperiodic variability between the rise and the decay of
the X-ray outbursts are found in this state: at the same
place in the color-color diagram, during the rise the source exhibits more
power at low frequencies ($\simless$ 1 Hz), whereas during the decay the
source is more variable at high frequencies ($\simmore$ 100 Hz). The
very-low frequency noise that characterizes the banana-state power spectra
below 1 Hz cannot be described in terms of a single power law but a
two-component model is required. In two outbursts a new 6-10~Hz QPO has
been discovered and tentatively identified with the normal/flaring
branch-like oscillation observed only at the highest inferred mass
accretion rates. We have compared the spectral and timing properties of
\aql\ with those of other atoll and Z sources.  Our results argue against a
unification scheme for these two types of neutron-star X-ray binaries.

\end{abstract}


\section{Introduction}

\aql\ belongs to the general group of systems known as low-mass X-ray
binaries (LMXB). These systems consist of a neutron star or a black hole
orbiting around a late-type star (later than A). The X rays are the result
of the accretion of material from the companion onto the compact star.
Mass transfer is thought to occur via Roche lobe overflow, and hence 
proceeds via an accretion disk. In \aql, the compact star is a neutron
star and the mass-donating companion is a V=21.6 K7V star, located at an
estimated distance of 2.5 kpc (Chevalier et al. 1999).  An orbital period
of 18.95 hours has been suggested (Chevalier \& Ilovaisky 1991; Welsh,
Robinson \& Young 2000).  The high-energy radiation is characterized by
1.5--2 month long transient X-ray outbursts during which the X-ray
luminosity can increase by more than three orders of magnitude. They are
thought to be due to thermal instabilities in the accretion disk (e.g. van
Paradijs 1996). Their recurrence time and duration vary but typical values
are $\sim$~200 days and 40--60 days, respectively ($\check{\rm S}$imon
2002). The source also displays Type I bursts (Zhang, Yu \& Zhang 1998)
that last a few tens of seconds and are interpreted as runaway
thermonuclear burning of matter on the surface of the neutron star. 


Various schemes have been proposed to categorize the LMXBs:  X-ray
spectral behavior as a function of intensity and the requirement of a
low-energy blackbody component in the X-ray spectra (Parsignault \&
Grindlay 1978; Naylor \& Podsiadlowski 1993), cluster analysis of a large
number of source characteristics (Ponman 1982), detailed X-ray spectral
fits (White \& Mason 1985), X-ray hardness-intensity and color-color
diagrams (Schulz, Hasinger \& Tr\"umper 1989), aperiodic variability at
very low frequencies (Reig, Papadakis \& Kylafis 2003).  Most relevant for
the purpose of this paper is the classification scheme in terms of the
rapid aperiodic variability and the patterns that these sources display
in  color-color diagrams (Hasinger \& van der Klis 1989). In this scheme
LMXBs are divided into two different subclasses, Z and atoll sources.

The three spectral branches that make up the Z shape are called horizontal
branch, normal branch and flaring branch and the two structures that occur
in atoll sources are known as the island and the banana (van der Klis
1989). At the lowest count rates an extension of the island state is
sometimes seen with a harder spectrum and stronger band-limited noise than
the "canonical" island state (van Straaten, van der Klis \& M\'endez 2003
and references therein). The term {\em extreme island state} has been used
to designate such state (Prins \& van der Klis 1997; Reig et al. 2000).
The spectral and timing properties of atoll sources in this state are
reminiscent of those seen in the low/hard state of black hole systems (van
der Klis 1994a, 1994b; Berger \& van der Klis 1998; Olive et al. 1998;
Belloni, Psaltis \& van der Klis 2002). Both Z and atoll sources move
through these patterns continuously, that is, without jumping from one
branch to another, although the source motion along the spectral tracks is
much slower in atoll sources when they are in the island state than in the
banana state and in Z sources in all states. The classification of Aquila
X--1 in this scheme was investigated by Reig et al. (2000) --- see also
Cui et al. (1998) ---, who studied the correlated X-ray timing and
spectral variations of \aql\ and presented evidence for its 
classification as an atoll source, exhibiting all classic atoll source
states.

This scheme has been revisited by Muno, Remillard \& Chakrabarty (2002)
and Gierli\'nski \& Done (2002). They reported that three transient atoll
sources which display a wide dynamic range in intensity ($F_{max}/F_{min}
\simmore 100$; 4U 1608--52, 4U 1705--44 and \aql\ trace out three-branch
patterns in the color-color diagram similar to those of Z sources, and
suggested this may be a general feature. Their study was based on the
spectral properties only. However, as pointed out by Hasinger \& van der
Klis (1989) it is difficult to make a clear distinction between Z and
atoll sources on the basis of the motion through the color-color diagram
alone. Analysis of the rapid aperiodic variability to study the noise
components in different regions of the color-color diagram and the actual
time scales of the motion of the source through the diagram are important
to identify source type and state. 

 In this work we have studied the spectral and timing properties of
\aql\ in order to investigate whether atoll and Z sources can be unified
into one classification scheme and provide new insights into the poorly
known extreme island states of atoll sources. We find that although the
color-color diagram shows a branch whose topology is similar to the normal
branch of Z sources, neither the timing properties nor the motion in the
color-color diagram of \aql\  agree with those of Z sources. 

\section{Observations}

\begin{figure}
\epsscale{1.5}
\plotone{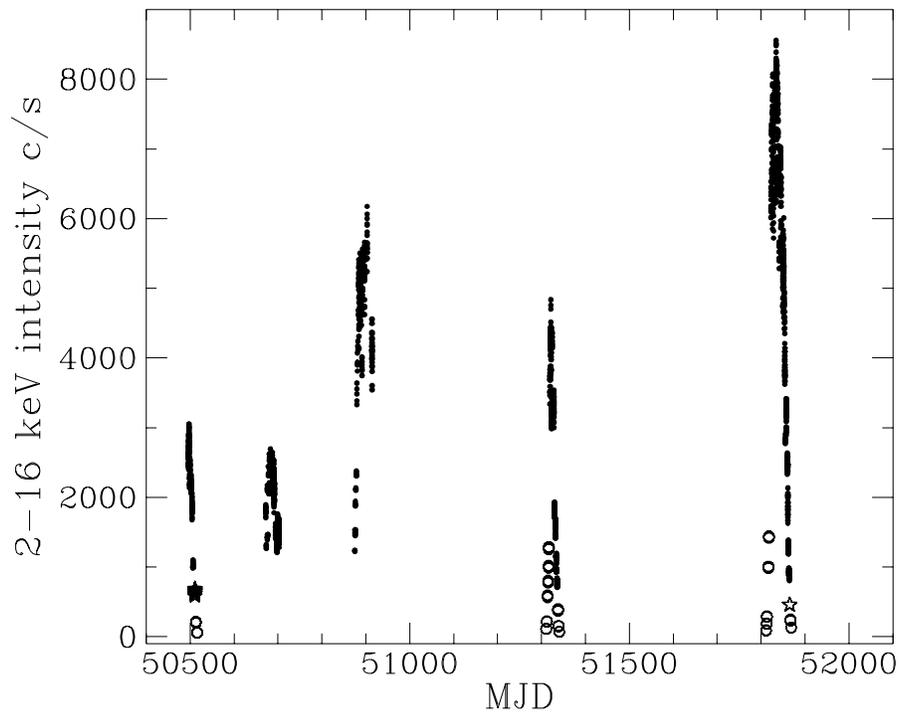}
\caption{Light curve of the entire set of observations showing
the five X-ray outbursts. Different symbols represent different 
spectral states: extreme island (circles), island (stars) and banana
(dots). 
\label{lightc}}
\end{figure}

The data were retrieved from the {\em RXTE} archive and comprise all
observations of \aql\ available from 1997 February to 2002 May.  Data
taken during satellite slews and Earth occultation were removed. Likewise,
all Type I bursts were excluded from our analysis.  The observations
contain five X-ray outbursts. Although the duration and maximum intensity
differ, the profile of the outbursts is very similar and is characterized
by a fast rise and a slower decay. During intensity maximum the light
curve is complex and multi-peaked. The fourth outburst showed two minor
outbursts (the peak intensity was one order of magnitude lower than the
main outburst) 70 days and 110 days after the main peak, respectively. 
The long-term light curve of the observations is shown in
Fig.~\ref{lightc}. 

\section{The color-color diagram}

\begin{figure}
\epsscale{1.5}
\plotone{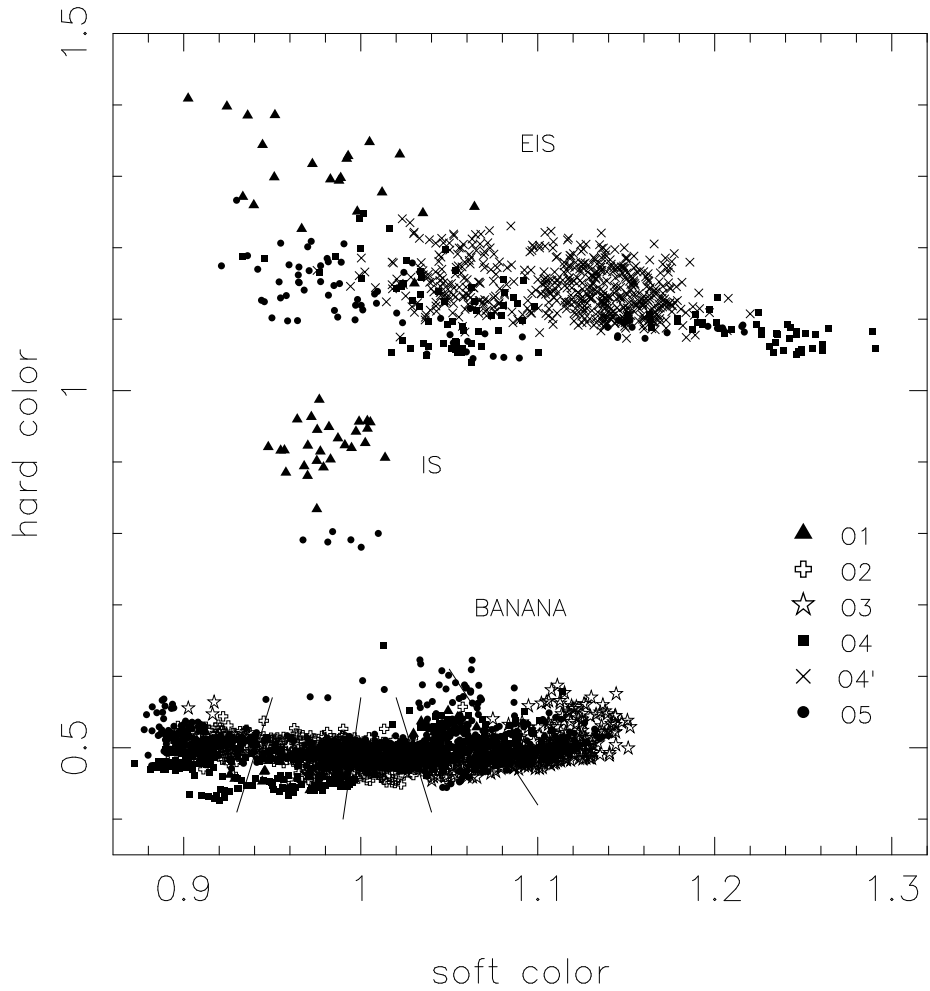}
\caption{Color-color diagram of \aql.  The soft and hard colors are
relative to those of the Crab. Each point in the color-color diagram 
represents 256 s. No error bars are given but all the points shown in this
plot have relative errors smaller than 5\%. Different symbols represent
different outbursts as indicated. The lines separate the banana-state
regions (from BS1 in the left to BS5 in the right).
\label{ccd}}
\end{figure}

Background subtracted light curves corresponding to the energy ranges
2.0--3.5 keV, 3.5--6.0 keV, 6.0--9.7 keV and 9.7--16.0 keV were used to
define the soft and hard colors as SC=3.5--6.0/2.0--3.5 and
HC=9.7--16.0/6.0--9.7, respectively.  The color-color diagram (CD) of
\aql\ was then constructed by plotting the hard color as function of the
soft color (Fig.~\ref{ccd}).  During the time spanned by the observations
the response of the detectors varied due to ageing.  In addition, gain
changes are applied occasionally, making the channel boundaries for a
given energy range change with time, and also slightly affecting the
effective areas of the detectors. Each gain change is the start of a new
"gain epoch". We reduced the effect of the color shifts that results from
these gain changes by linearly interpolating between the count rates in
the two energy channels straddling the energy boundaries in each epoch.
Each data point of \aql\ was normalized with the closest corresponding
Crab point within the same gain epoch in order to mitigate the response
change effects on the colors. The final CD was obtained by averaging the
five (one for each detector) normalized CDs and rebinning into 256-s data
points. We excluded data for which the resulting relative errors were
larger than 5\%. The total number of data points excluded from the
analysis was $\simless$ 10\%. In order to recover the true values of the
colors of \aql\ the soft color should be multiplied by 2.35 and the hard
color by 0.56 (quoted values are averages of the Crab colors during the
observations and for all five PCUs). The variation of the Crab colors
computed as the root-mean-square, i.e. (the standard deviation over the
mean color) was 5.3\% and 1.7\% for the soft and hard colors,
respectively --- see also Fig.~1 in van Straaten et al. (2003).

\subsection{Spectral states}

\aql\ can be found in two main states: a low-intensity hard state  and a
high-intensity soft state. In addition, a short-lived intermediate or
transition state is found displaying values of the hard color in between
the two main states.  The aperiodic variability as a function of spectral
hardness, i.e. position in the CD of \aql\ was previously studied by Reig
et al. (2000). With only two of the five outbursts analysed in that work,
the CD consisted of a soft branch, the banana branch, and two isolated
groups of points which were called extreme  island and island states. The
large amount of data now analyzed reveals a more structured CD, in which
the extreme island state appears as a more stable and longer-lived branch
than the island state.  The high/soft state corresponds to the classical
banana state and the island state represents the transition between the
two main states. 


Outbursts 1, 4 and 5 (O1, O4 and O5) contain points in the two
main states while outbursts 2 and 3 (O2 and O3) provide points to the
banana state only. Observations of O2 and O3 began and finished when
the source was still at a relatively high level of emission ($> 1200$ c/s)
and did not experience large amplitude changes (I$_{max}$/I$_{min}$ $<$
5). In contrast, outburst 1, 4 and 5 extended over a larger dynamic range
in intensity (I$_{max}$/I$_{min}$ $>$ 50). None of the points of the two
minor outbursts which followed O4 (which will be termed here as
O4$^\prime$) contributed to the banana state. 

Outbursts 1, 4, 4$^\prime$ and 5 include points in the extreme island
state. While O4 and O5 include points both during the rise and during the
decay, O1 gives points during the decay only. The island state  occurred
only during the decay of O1 and O5. All five outbursts (except for
O4$^{\prime}$) have points in the banana state but only O3 provides banana
state points during the rise.

\subsection{Motion in the color-color diagram}

\begin{figure}
\epsscale{1.5}
\plotone{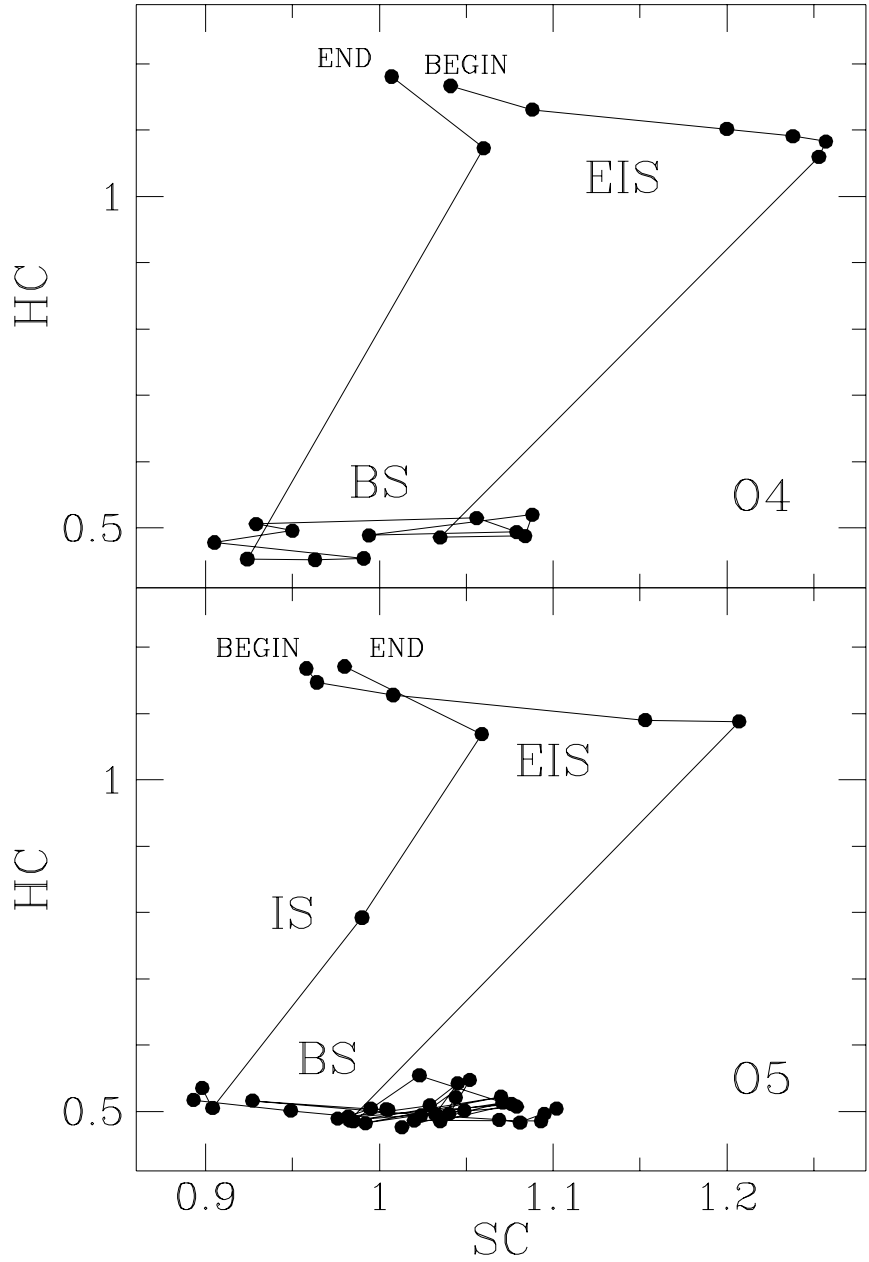}
\caption{Track followed by the source in the color-color diagram during
outbursts 4 and 5. Each data point is a one day average.
\label{CDmotion}}
\end{figure}

\begin{figure}
\epsscale{1.5}
\plotone{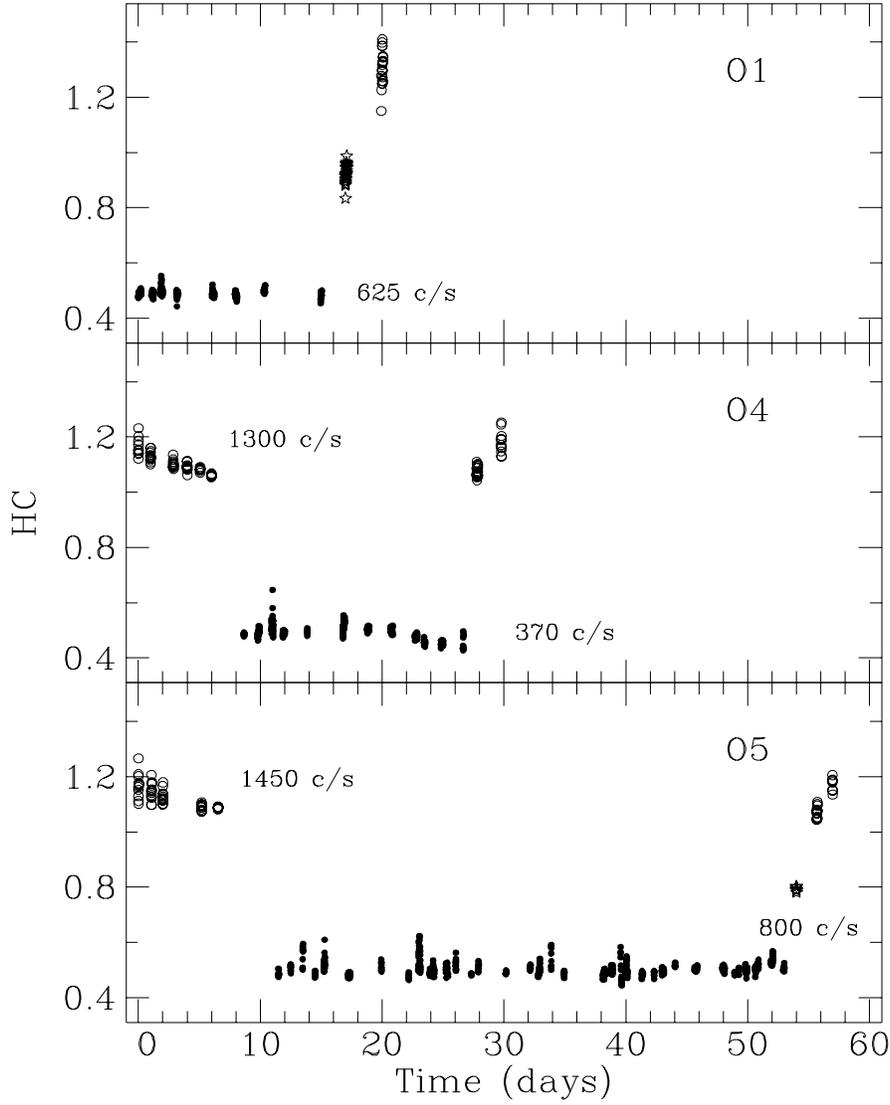}
\caption{Evolution of the hard color showing the
transitions between the different spectral states: extreme island (circles), 
island (stars) and banana (dots). Time refers to the onset
of each outburst. 
\label{coltime}}
\end{figure}

Figure \ref{CDmotion} displays the motion of \aql\ in the CD for outbursts
4 and 5. Data points are $\sim$1-day averages. Figure~\ref{coltime} shows
the evolution of the colors with time for those outbursts that include
spectral transitions (O1, O4, O5). The 2--16 keV intensity just before the
transitions is also given. All quoted intensities in this section are
background subtracted and correspond to 5 PCUs in the energy range 2--16
keV. There is a strong correlation between the position of the source in
the CD and its intensity. At the onset of the outbursts the data points
distribute in the softest part of the extreme island state. The count rate
is $\simless$  100 c/s. As the intensity increases the source moves toward
the right along the extreme island branch. The hard color (HC) slightly
decreases on average. In about 6 days the count rate increases by one
order of magnitude and the soft color (SC) increases by $\approx$+0.3.
Then the source seems to jump to the banana state, recovering the initial
values of the soft color. The hard color decreases by $\approx -0.5$. The
2--16 keV intensity is $\sim$ 3 or 4 times that of the extreme island
state prior to the spectral transition, namely 4000--6000 c/s. No
intermediate points between the two states are observed during the rise of
the outbursts. This can be attributed to the fast rise and the lack of
good sampling of the data. Indeed, the observational gaps between the last
point of the extreme island branch and the first one of the banana branch
were 2.6 and 3.1 days for O4 and O5, respectively. 

\begin{figure}
\epsscale{1.5}
\plotone{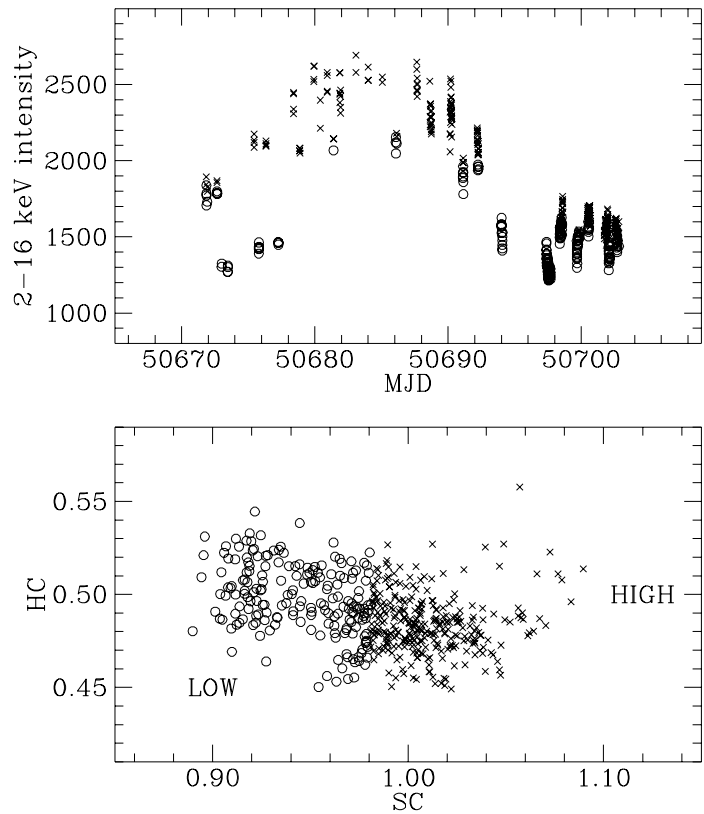}
\caption{Light curve of the second X-ray outburst (upper panel) and the 
corresponding color-color diagram. The terms {\em high} and {\em low} refer
to the high-intensity and low-intensity points of the flares.
\label{flare2ccd}}
\end{figure}

As the intensity continues to increase the source becomes harder, i.e., it
moves to the right along the banana, with approximately constant hard
color. The peak of the outburst is characterized by flaring activity with
erratic changes in count rate. The count rate at the peak is 50--100 times
the minimum detected count rate.  Despite this irregular behavior the
correlation between the intensity and the colors is maintained in the
sense that the lower intensity points in the flares display a lower soft
color as it is illustrated in Fig.~\ref{flare2ccd}.  In other words, 
during the flare maxima the source lies at the right end of the banana
branch, and in between flares it moves to the left. Thus the flaring
variability in the light curve translates into the CD in a back and forth
motion which approximately extends over the right half part of the banana
state. This motion is quite fast. As an example, in one of the flares in
O5 the soft color changed by 0.08 in about 2.4 hr. In another flare by
0.06 in 0.55 hr. This flaring behavior is seen in all outbursts but O2
shows it exclusively (Fig.~\ref{flare2ccd}). Such behavior stops once the
source gets to half way the decay of the outburst, at which point the
source shifts to lower soft colors (to the left along the banana branch). 

As the outburst declines the source moves back along the banana branch,
abandoning it when the count rate becomes lower than $\sim$ 800 c/s. The
transition to the hard state occurs at much lower soft color than when it
entered the banana, although not necesarily at the lowest soft color.  The
banana state covers values of the soft color between 0.87 (O5) and 1.16
(O3). The transition to the hard state takes place at SC $\approx$
0.92--0.93 (Fig.~\ref{CDmotion}).  The count rate at which the transition
between the two main states takes place is higher when the source is in
the rise of the outburst. Hard to soft transitions occur when the count
rate is well above $\sim$ 1000 c/s. Soft to hard transitions when the
count rate is well below $\sim$ 1000 c/s as it can be seen in
Fig.~\ref{coltime}.

Rather than jumping directly to the hard branch the source remains for a
short time ($\le 0.1$ days) in an intermediate state, the island state.
Note that such a short island state episode would usually have been missed
due to data gaps in the rise. The count rate in the island state is $\sim$
200--400 c/s. At even lower count rates the source finds itself back in
the extreme island branch, moving toward the left as the intensity
decreases, to eventually become too weak to measure the colors
sufficiently well.  The source re-enters this state during the decay when
the count rate goes below $\sim 200$ c/s. The re-entry point occurs at a
lower soft color (SC $\approx$ 1.05--1.10) then when the source left the
extreme island state (SC $\approx$ 1.3). As the speed of motion along the
extreme island state is approximately constant, the time during which the
source can be found in this state is shorter during the decay of the
outburst than during its rise. A difference in the position of the source
in the CD depending on whether the source intensity increases or decreases
was already recognized by Reig et al. (2000). 

The main result that should be stressed is that transitions between states
do not occur at the same point of the spectral branches. During the rise
of the X-ray outburst the source occupies the hardest parts of the extreme
island state before the spectral transition to the banana state. During
the decay it tends to occupy the softest part of the banana branch before
moving into the island state. However, the points of departure and arrival
are different. In the CD, this translates into some sort of rectangular
track (Fig.~\ref{CDmotion}) along which \aql\  moves clockwise as the
count rate first increases and then decreases.

In addition to Fig.~\ref{coltime}, Tables \ref{dur} and \ref{timetran}
also illustrate the time scale for the motion through the diagram.
Table~\ref{dur} and Table~\ref{timetran} give upper limits on the duration
of the source in each state and the time scales of the spectral
transitions, respectively. Note that the observational gap for the
spectral transitions from the extreme island state to the banana state is
approximately two times longer than viceversa. Given the speed of motion
of the soft color along the extreme island branch (roughly 0.05
day$^{-1}$) a three-day gap might imply a change in soft color of about 0.15.
Thus, it is possible that the actual entry point into the banana branch is
at lower values of the soft color. In this case  the motion of \aql\ in the CD
would be similar to that of 4U 1705-44 (Barret \& Olive 2002), in which 
both the transition from (to) the extreme island state to (from) the
banana state occur at the same point of the banana branch. The motion in
the CD then would then resemble an inverted triangle rather than a
rectangle.


\section{Timing analysis}

In order to investigate the aperiodic variability of \aql\ we obtained
power spectra by dividing the 2--60 keV PCA data of each observation into
256-s segments and calculated the Fourier power spectrum of each segment
up to a Nyquist frequency of 2048 Hz. The  high-frequency end (1500--2048
Hz) of the power spectra was used to determine the underlying Poisson
noise (approximated by a constant power level), which was subtracted
before performing the spectral fitting.  
Our aim is to investigate the aperiodic variability in correlation with
the spectral states, which in turn, correlate with the source count rate.
Thus we divided the color-color plane into various regions and obtained a
mean power spectrum for each region. The mean power spectrum was the
result of averaging all the power spectra corresponding to the points
enclosed in each color region. See Table~\ref{reg} to find the average
values of the colors, boundaries and intensity of each region. Prior to
this, we investigated whether power spectra from the same CD region taken
at different outbursts were consistent with having the same properties. We
found that while this was true for the banana state, some differences
existed in the extreme island state depending on whether the source was in
the rise or the decay of the X-ray outburst.

The extreme island state is populated with points from the rising part of
O4 and O5, the decaying part of O1, O4 and O5 and from the extended
extreme island state of O4$^{\prime}$. Consequently, the extreme island
state was first separated into points pertaining to the rise, to the decay
or to O4$^{\prime}$. Then each one of these groups was further divided
into three subgroups according to the value of the soft color. Note that
in the case of the rise and decay regions this division implies a division
in count rate as well as in hard color. On average, the extreme island
sates EISrise1 and EISdecay2,
(see Table~\ref{reg}) have lower intensity and are softer than EISrise2 and
EISdecay1, respectively. In turn, EISrise2 contains softer points than
EISrise3. The island state points (intermediate state) defined another
region. The timing properties of the banana state in atoll sources have
been seen to vary smoothly as the soft color, or equivalently, the count
rate increases, that is, from the lower (left) to the upper (right) banana
branch. Therefore, the banana branch was split into five regions (BS1 to
BS5) in increasing order of the soft color. Each region approximately
contains the same amount of data.

There is no unanimity in the literature about the names of the power
spectral noise components nor about the mathematical functions used to fit
those components. In this work  we have followed the terminology of
Belloni et al. (2002) and fitted the power spectra using Lorentzian
functions only. Also, for  displaying our power spectra we have used the
power times frequency representation $\nu \times P(\nu)$ vs $\nu$, in
which the product of the power density in the rms normalization and the
Fourier frequency is plotted as a function of the Fourier frequency
(Belloni et al. 1997). As fit parameters we use the frequency  $\nu_{\rm
max}$ at which  the Lorentzian attains its maximum value in such 
representation, and the $Q$ value, which gives a measure of the coherence
of the Lorentzian. In terms of the half width at half maximum or $\Delta$,
these quanties are related by $\nu_{\rm max}=\sqrt{\nu_0^2+\Delta^2}$ and
$Q=\nu_0/2\Delta$, where $\nu_0$ is the central frequency of the
Lorentzian. Between three and five Lorentzians were needed in order to get
a good fit. If during the fitting procedure a Lorentzian component
resulted with negative $Q$ value then we set it to zero.

It should be noted that the naming scheme of Belloni et al. (2002) was
applied, in the case of low-luminosity neutron stars, to systems
displaying the island state only. Thus,  a complete description of some
power spectra, especially those in the banana state, in terms of the noise
components given in that work is not always possible and extra or new
components are needed in order to obtain acceptable fits. The extensions
to the Belloni et al. naming scheme we use are those of van Straaten et
al. (2003).

In the terminology of Belloni et al. (2002), the band-limited noise of
atoll sources in the (extreme) island state consists of three Lorentzians
namely, $L_b$ that accounts for the power spectrum at low frequencies
($\simless 1$ Hz), $L_\ell$ and $L_u$ that fit its high-frequency end. 
$L_\ell$ may show up as a broad bump at frequencies 10--25 Hz (normally in
the island state) or as a narrow QPO at frequencies $\simmore$ 100 Hz
(normally in the banana state). In fact, at frequencies above $\simmore$
100 Hz $L_\ell$ and $L_u$ represent the kHz QPOs seen in low-mass X-ray
binaries. However, note that this identification of $L_\ell$ with both a
broad peak and the lower kHz QPO is only tentative and the broad bump at
10-25 Hz might alternatively represent a separate component (van Straaten
et al. 2002). The intermediate frequency range (1--100 Hz) is described by
one broad Lorentzian or "hump", $L_h$ and one narrow Lorentzian $L_{LF}$
referred to as the low-frequency QPO. Not detected by Belloni et al.
(2002) but also present in some atoll sources (van Straaten et al. 2002)
is the so-called hectoHerz QPO, $L_{hHz}$, a broad feature with
characteristic frequency at $\sim$ 100 Hz.

\subsection{Spectral states: aperiodic variability}

\begin{figure*}
\epsscale{0.60}
\begin{tabular}{cccc}
\plotone{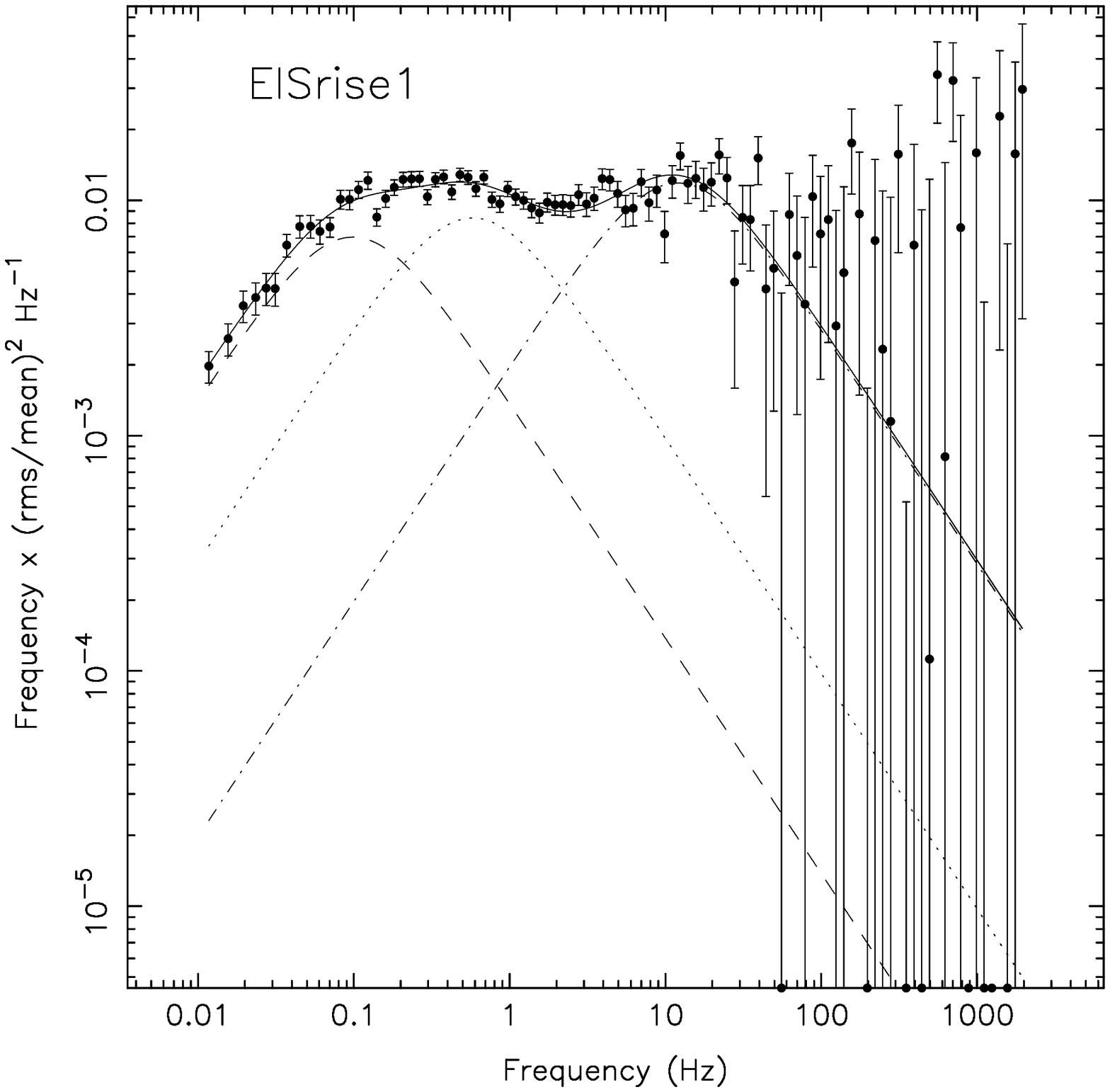} &
\plotone{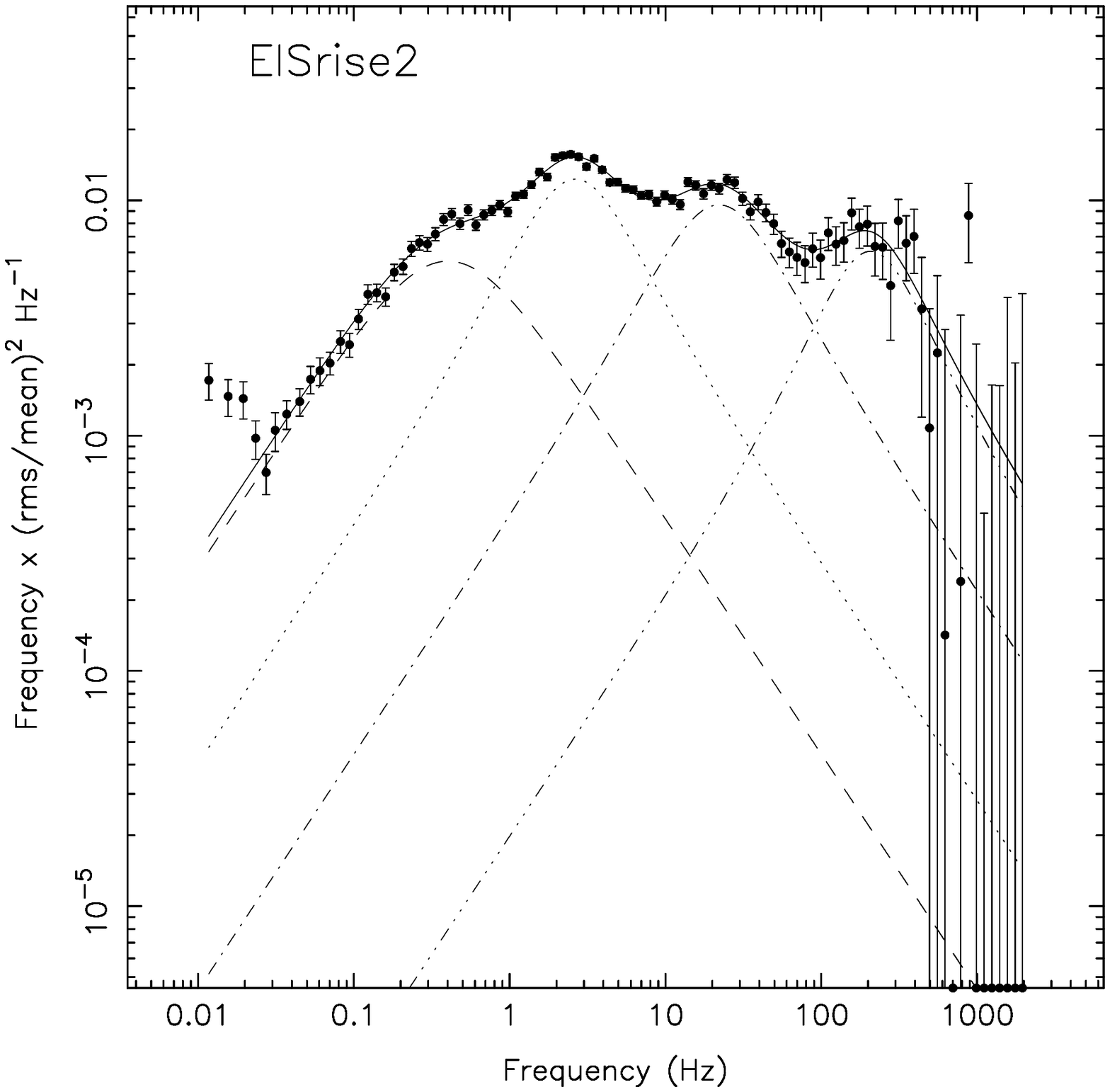} &
\plotone{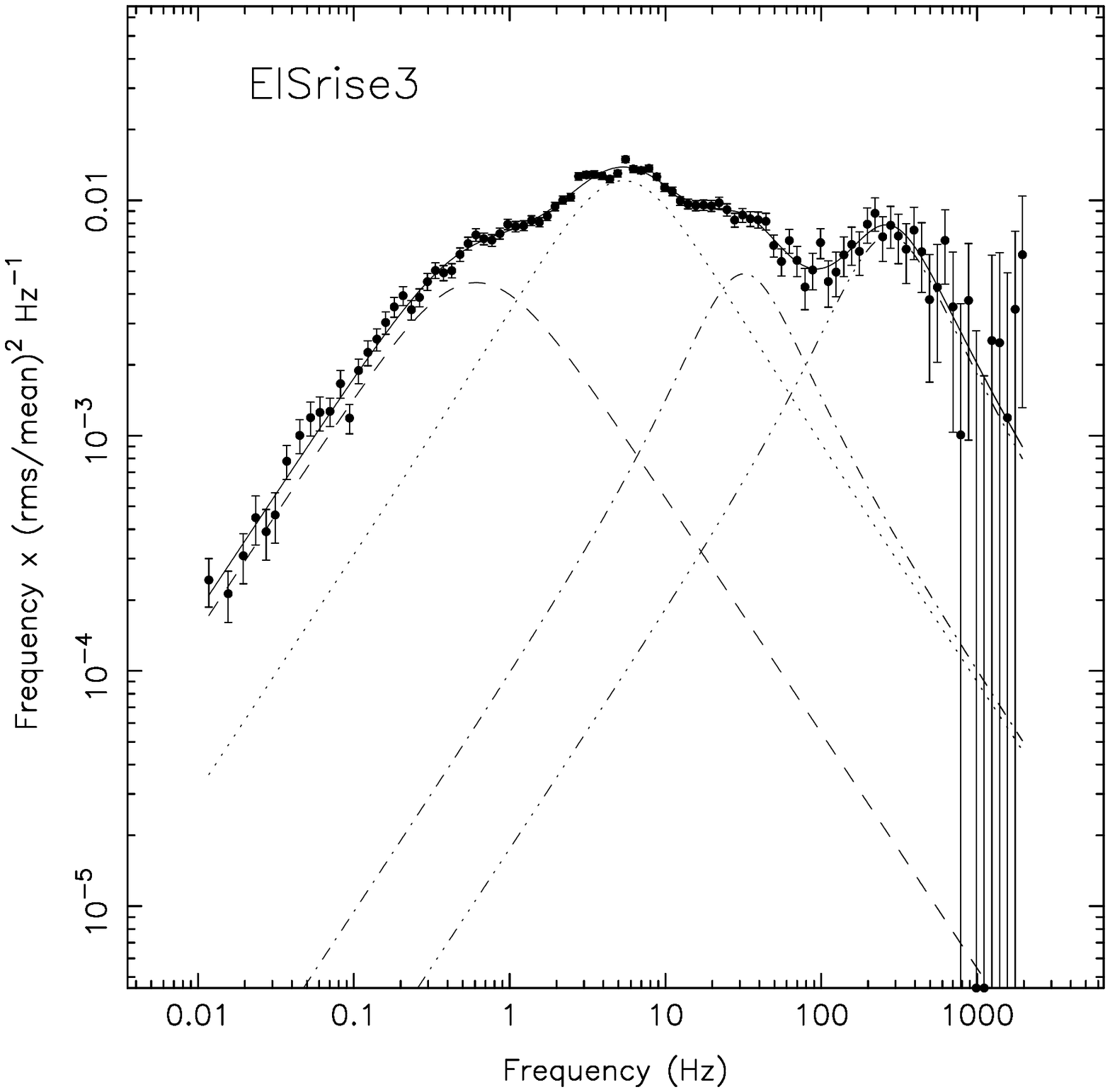} & \\
\plotone{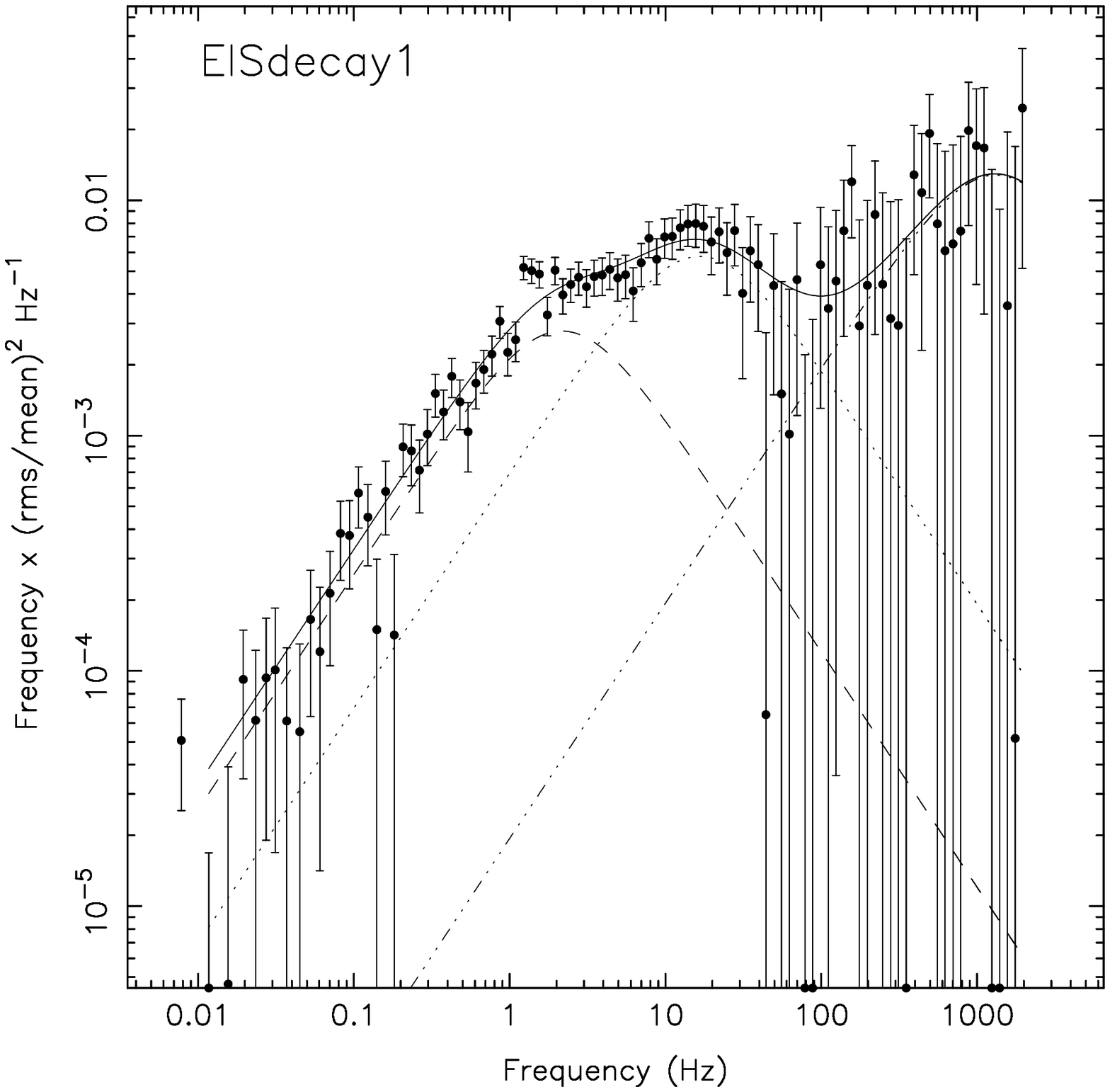} &
 \plotone{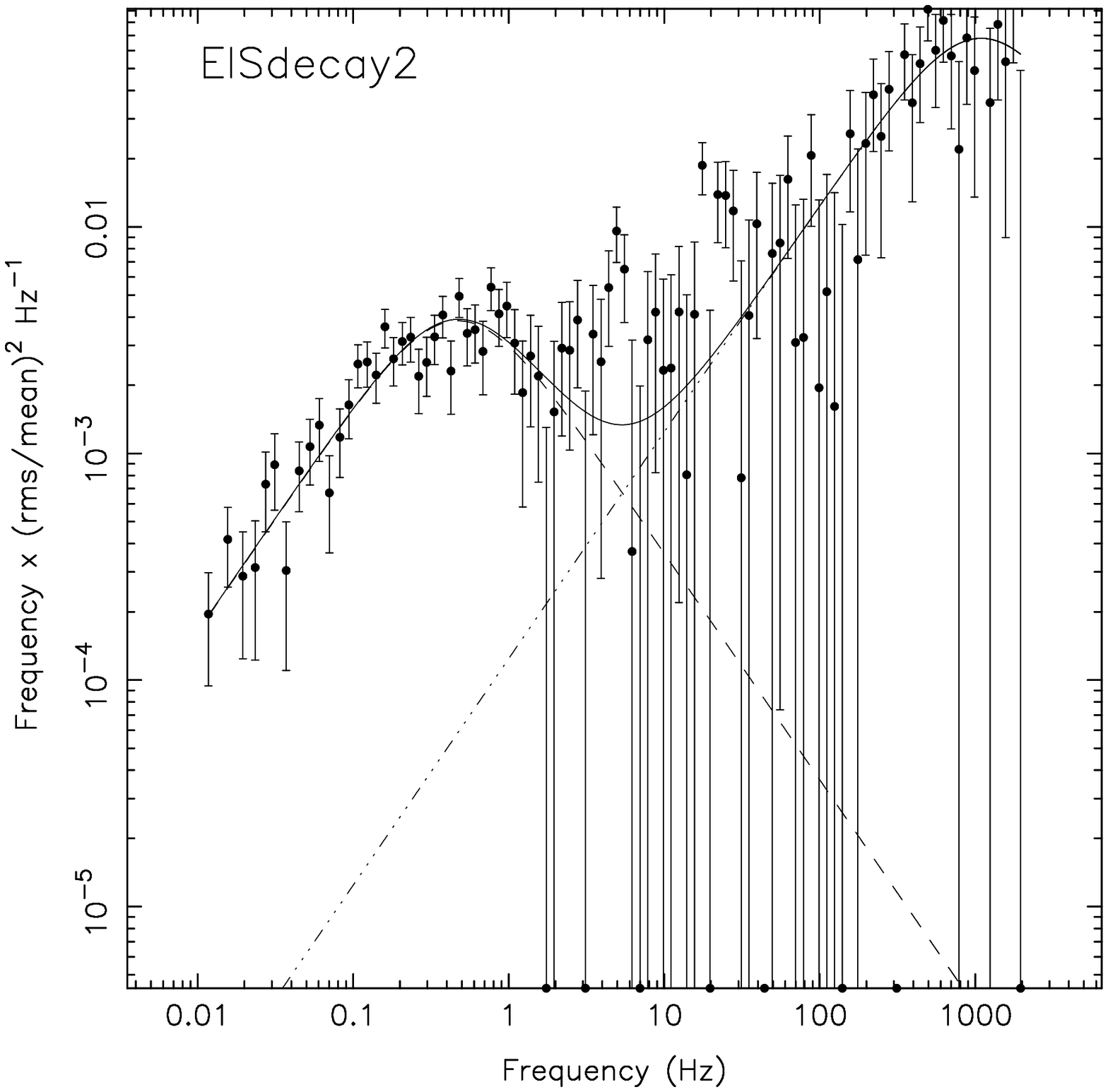} &
\plotone{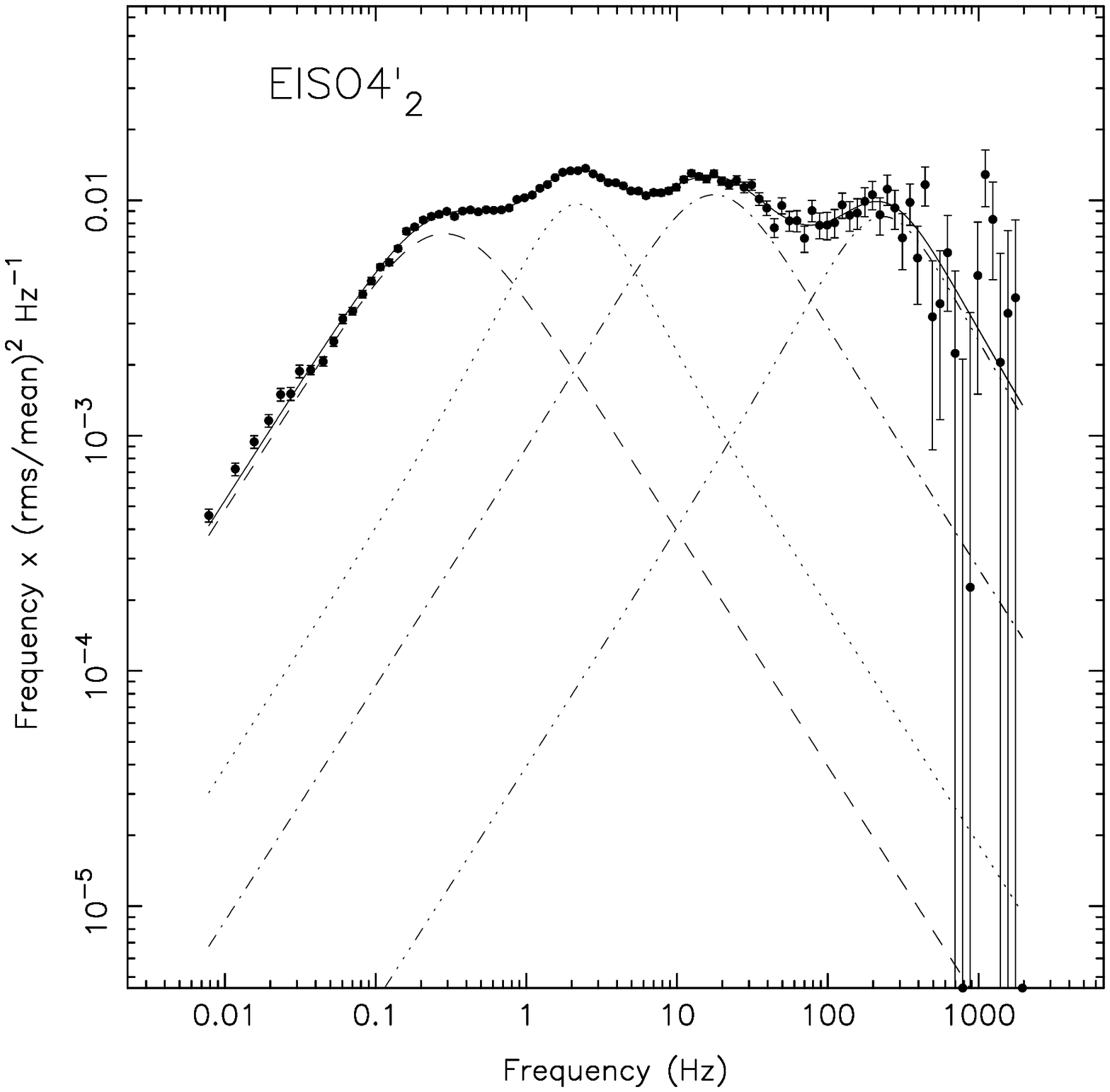} & \\
\plotone{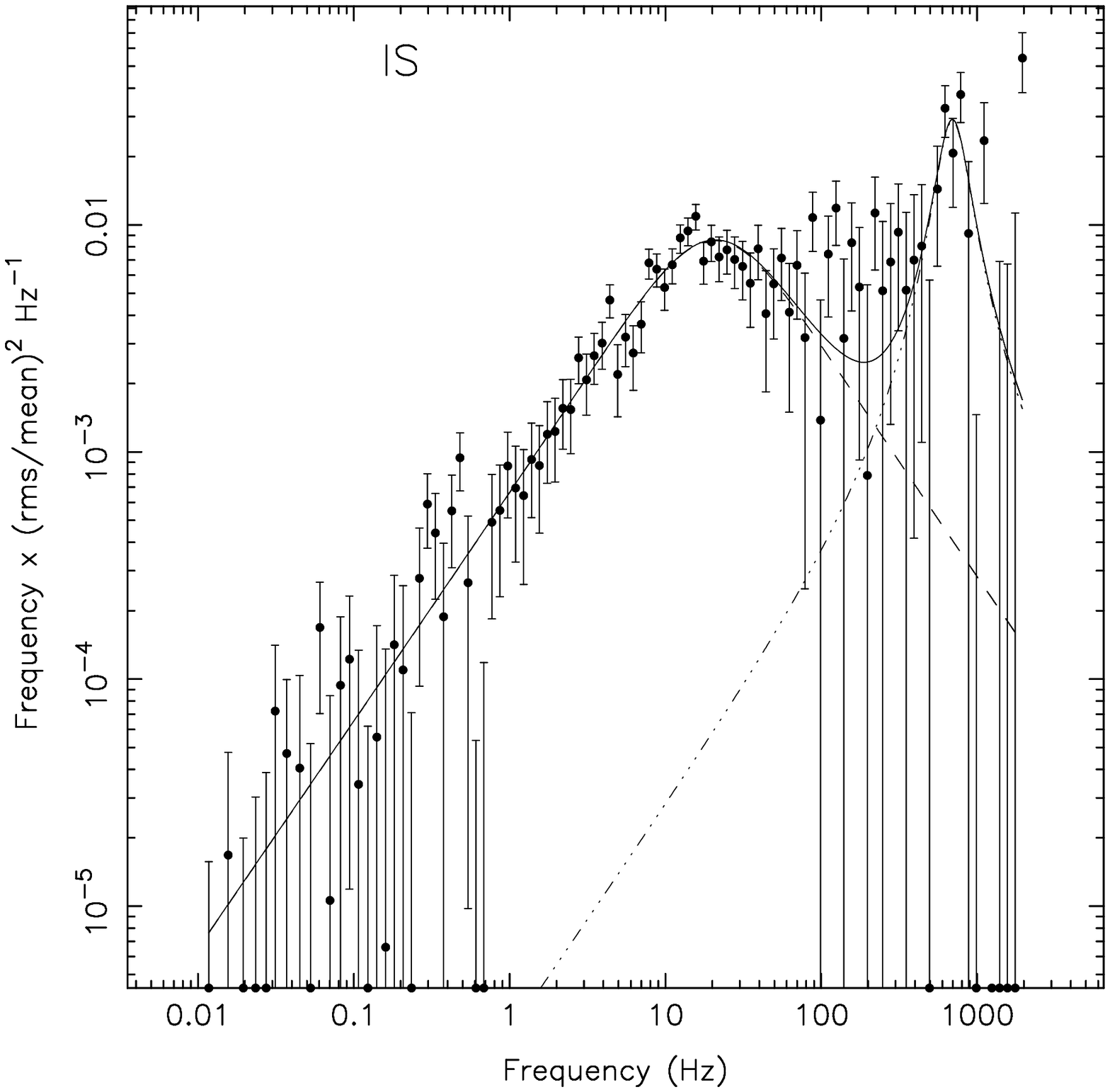} & 
\plotone{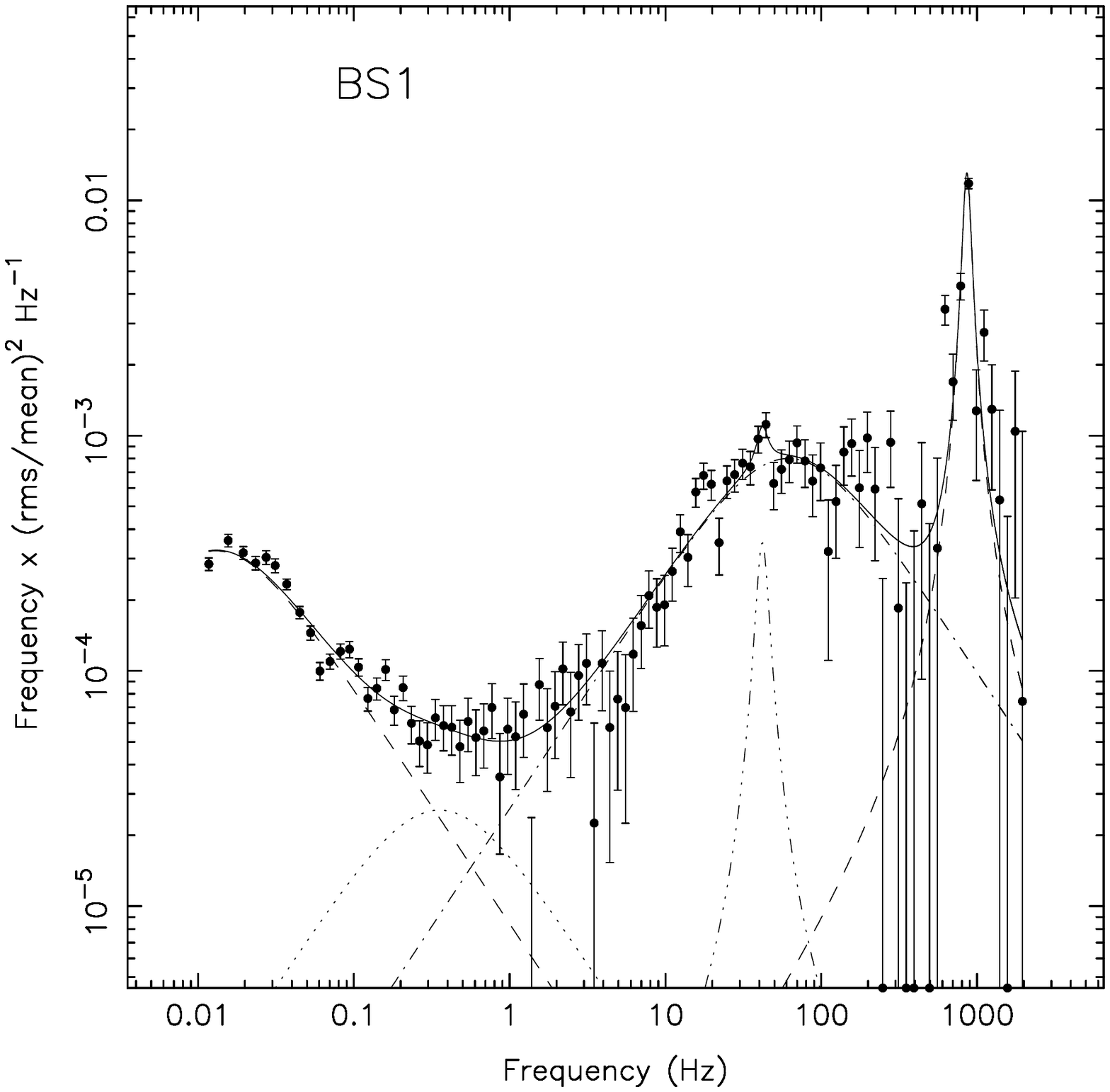} &
\plotone{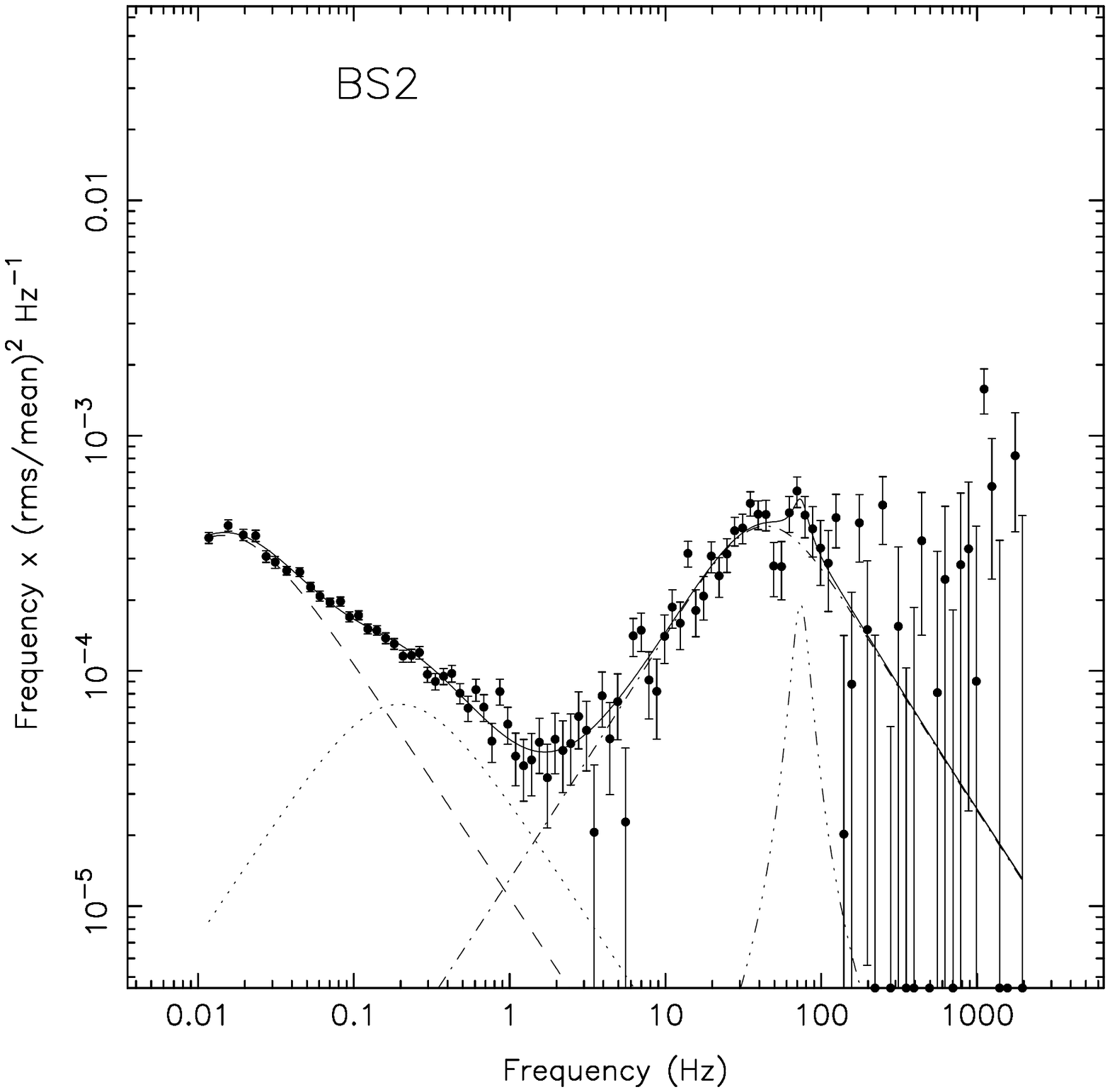} & \\
\plotone{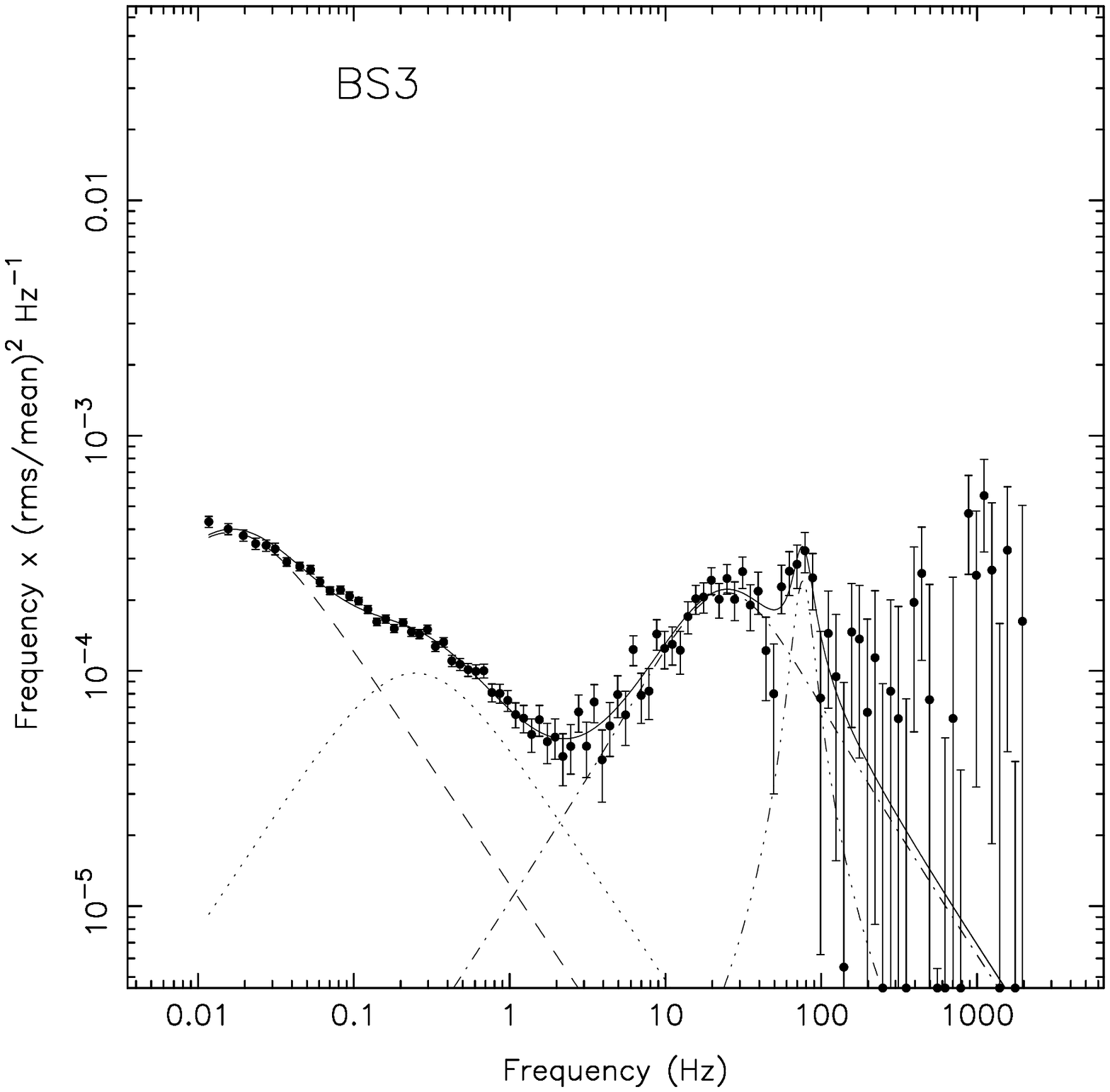} &  
\plotone{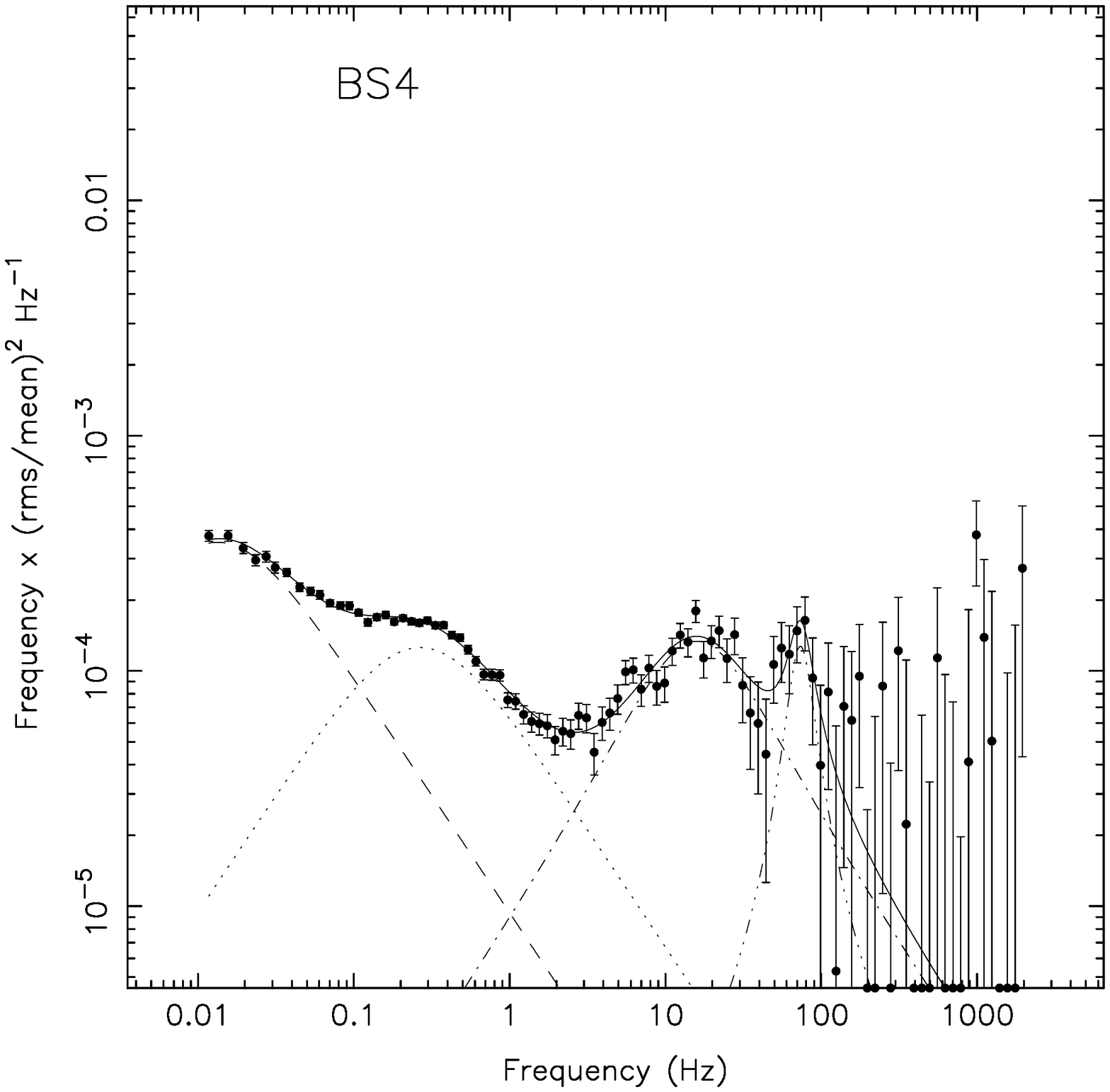} &
\plotone{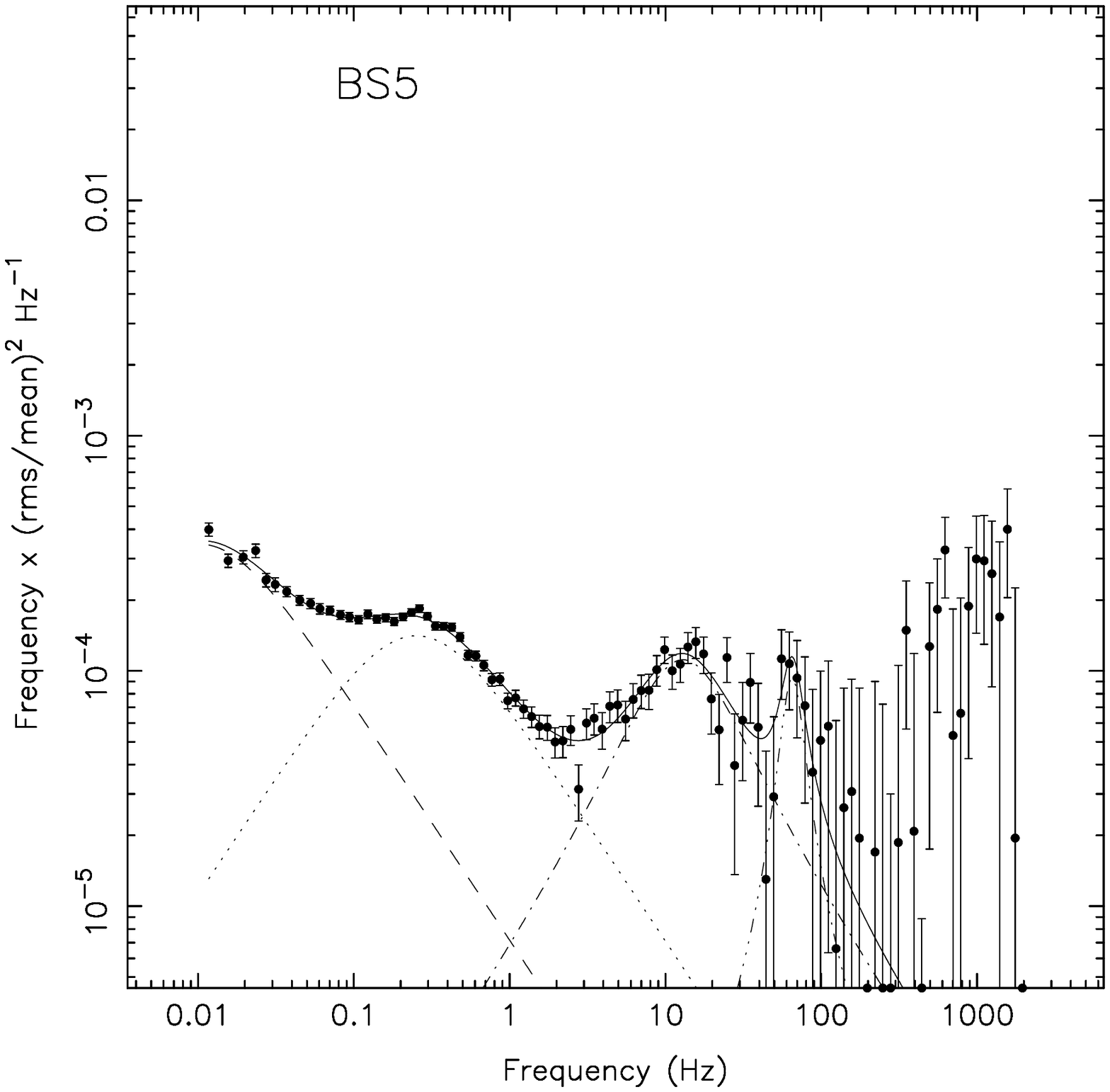} & 
\end{tabular}
\caption{Power spectra and best fits for different positions in the
color-color diagram (see Table~\ref{reg}) The lines represent the different
noise components as follows: for the island states, $L_b$ (dashed),
$L_h$ (dotted), $L_\ell$ (dot-dashed) and $L_u$ (dot-dot-dot-dashed); for
the banana state $L_{LVLFN}$ (dashed), $L_{HVLFN}$ (dotted), $L_b$
(dot-dashed), $L_b{_2}$ (dot-dot-dot-dashed). The second dashed line in BS1
corresond to $L_\ell$.
\label{pds}}
\end{figure*}

 Table \ref{fitres} lists the results of the fits and Fig.~\ref{pds}
depicts the power spectra and fit functions for different regions of the
CD (see Table~\ref{reg} to identify the position in the CD).

\subsubsection{The extreme island state (EIS)}
\label{EIS}

The extreme island state power spectrum  contains, in order of increasing
frequency, the low-frequency band-limited noise, $L_b$, the broad peak
low-frequency noise $L_h$, the broad version of $L_\ell$ and one more
component which could be either the hHz QPO or a kHz QPO. Based on the
rms/frequency correlations of the noise components seen in other atoll
sources, van Straaten et al. (2002, 2003) favored the upper kHz
interpretation. If $L_\ell$ is also interpreted as the lower kHz QPO then
the  intervals EISrise2, EISrise3 and ISO4$^{\prime}_2$  (see
Fig.\ref{pds}) would represent the first detection of the two kHz QPO in
\aql. Alternatively, the fourth component could be the hHz QPO. Note that
the narrow $\sim$1160 Hz peak in EISO4$^{\prime}_2$ is not statistically
significant. An F-test reveals that the probability that the improvement
of the fit (in terms of a lower $\chi^2$) by the addition of an extra
Lorentzian happen by chance is larger than 30\%. The characteristic
frequencies of the noise components increase and their rms values decrease
slightly as the count rate increases, i.e., the source moves toward the
right in the extreme island state. 

\begin{figure}
\epsscale{0.75}
\begin{tabular}{ccc}
\plotone{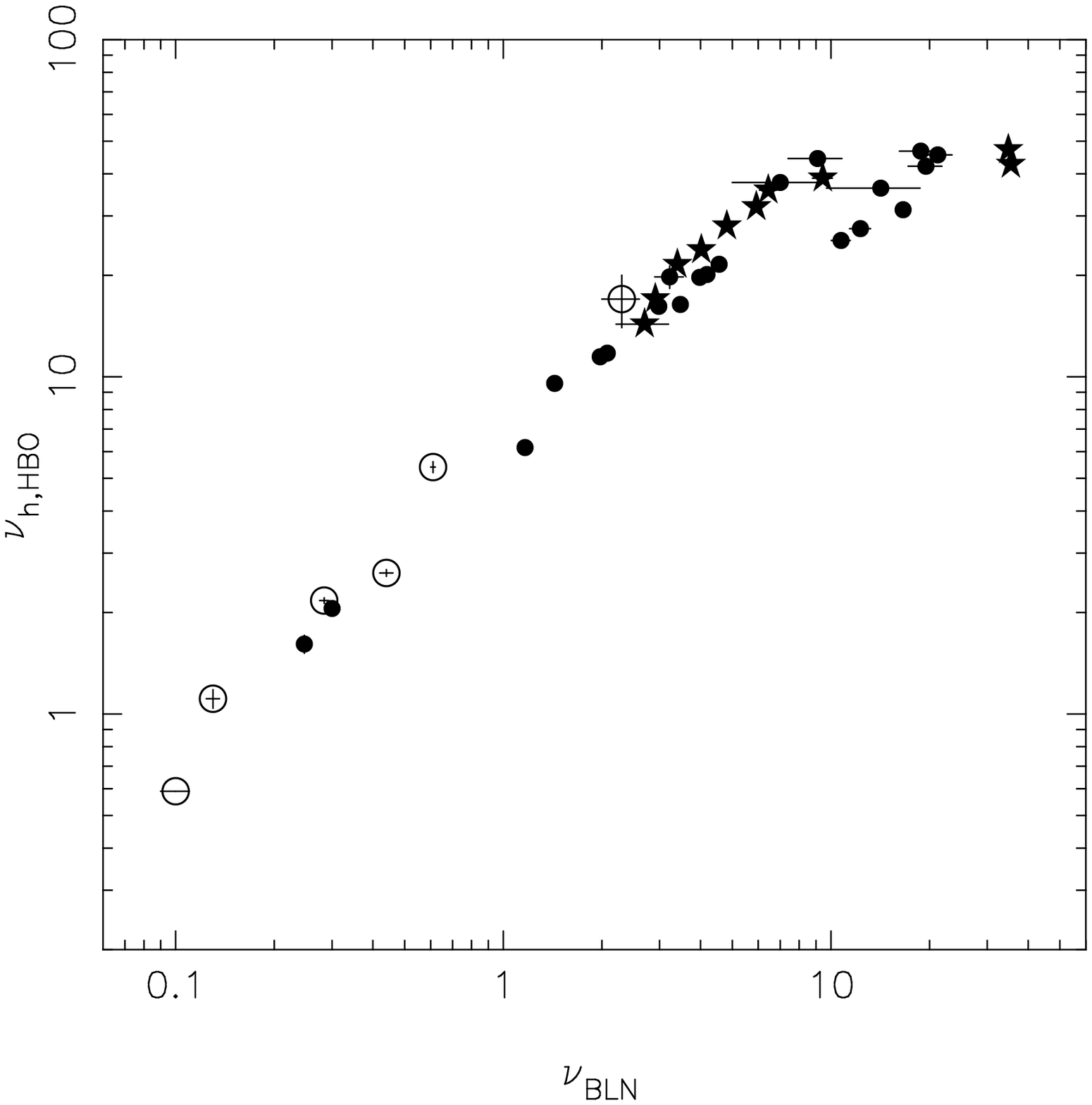} &
\plotone{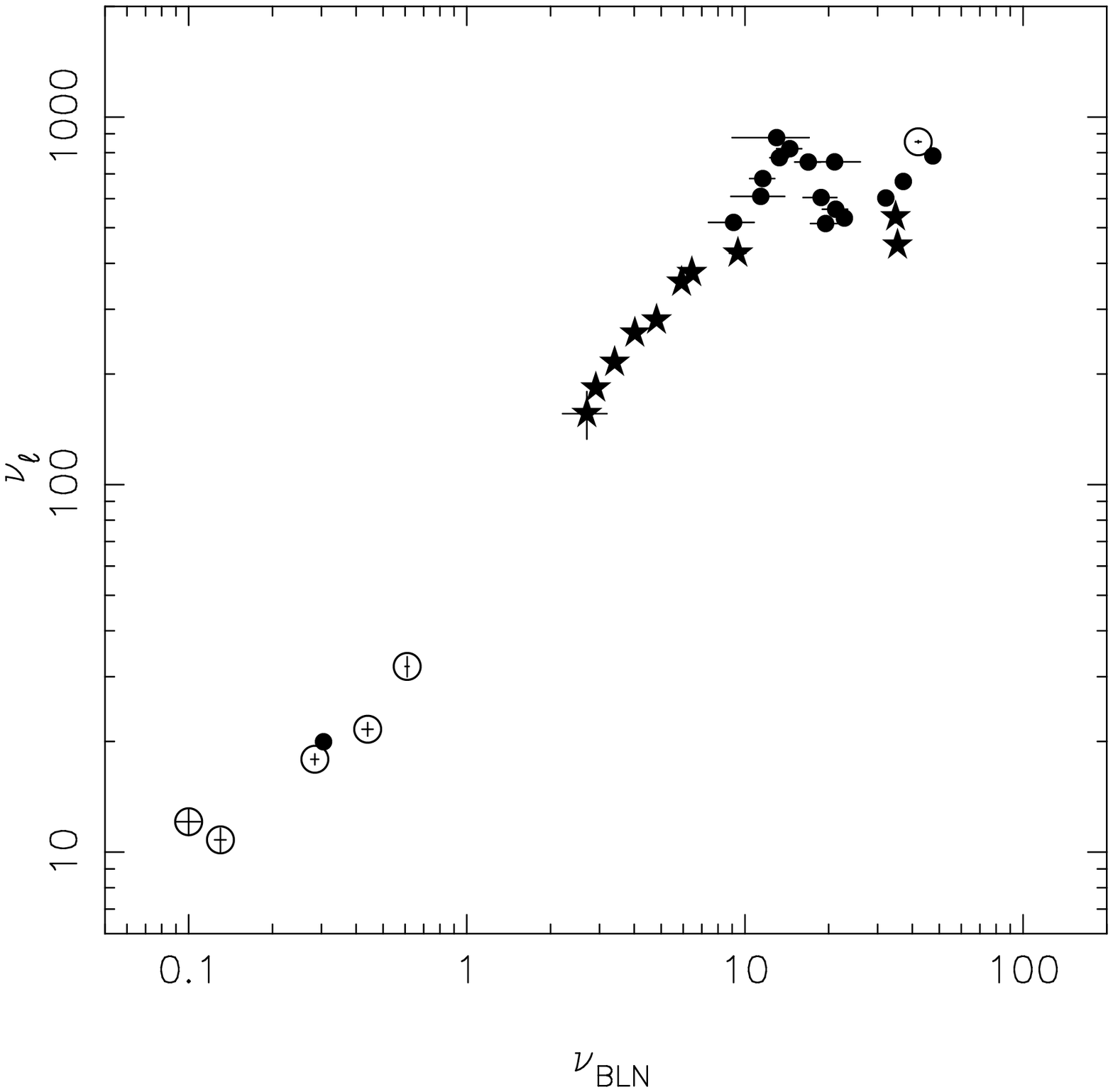} & \\
\end{tabular}
\caption{Characteristic frequency of the $L_h$ and $L_\ell$ noise components 
of the atoll sources 4U 1608--52, 4U 1728--34 and 
4U 0614+09 (dots) and the HBO of the Z sources GX 5-1 (stars) as a function 
of the characteristic frequency of the band-limited noise ($L_b$ for the atoll
sources and HBO for GX5--1). 
\aql\ observations have been represented by open circles.
\label{corr}}
\end{figure}

Figure~\ref{corr} displays the characteristic frequency of the  $L_h$ and
$L_\ell$ components as a function of the band-limited noise component
$L_b$ for \aql, the atoll sources 4U 1608--52 (van Straaten et al. 2003),
4U 1728--34 and 4U 0614+09 (van Straaten et al. 2002) and the Z source
GX5--1 (Jonker et al. 2002). In the case of GX5--1 te horizontal branch
oscillation is plotted instead of $L_h$. Note also that the broad-band
noise in the Z source was not fitted with a Lorentzian but with a cutoff
power law, $P(\nu)\propto \nu^{-\alpha}\, e^{-\nu/\nu_{cut}}$. By
differentiating $\nu P(\nu)$ and equating the resulting function to zero
the maximum frequency of the Z band-limited noise,
$\nu_{max}=(1-\alpha)\nu_{cut}$, was derived.  Figure~\ref{corr} seems to
confirm that {\em i)} the 1--20 Hz broad feature $L_h$ observed in atoll
sources at low count rates (island states) has the same origin as  the
horizontal branch oscillations (HBO) observed in Z sources (Psaltis,
Belloni \& van der Klis 1999, van Straaten et~al. 2003), {\em ii)} the 
broad bump at frequencies 10--25 Hz is associated with the lower kHz QPO.
The extreme island state data in \aql\ sample the lowest characteristic
frequencies so far detected. 

\begin{figure}
\epsscale{0.75}
\begin{tabular}{ccc}
\plotone{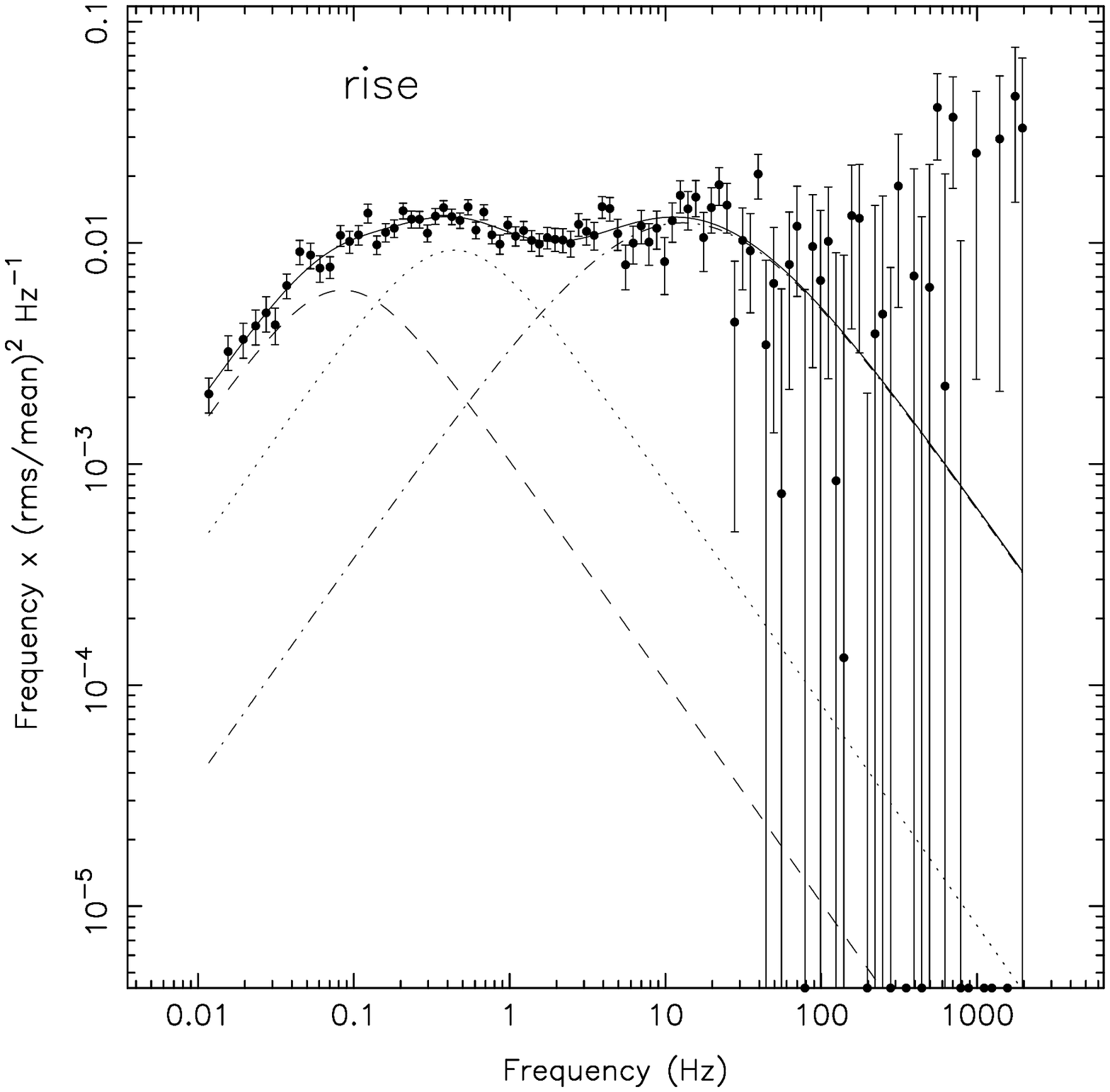} &
\plotone{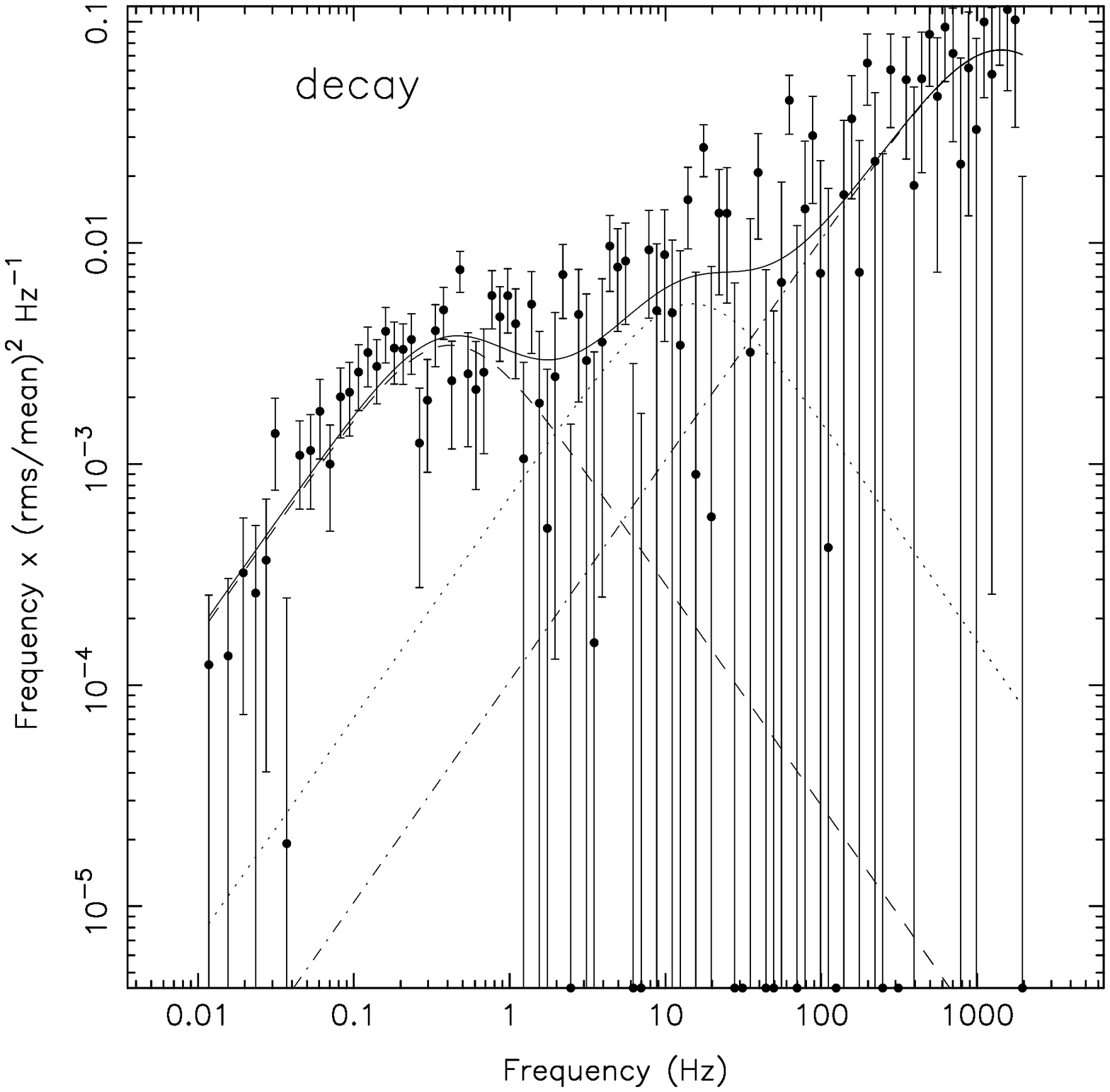} & \\
\plotone{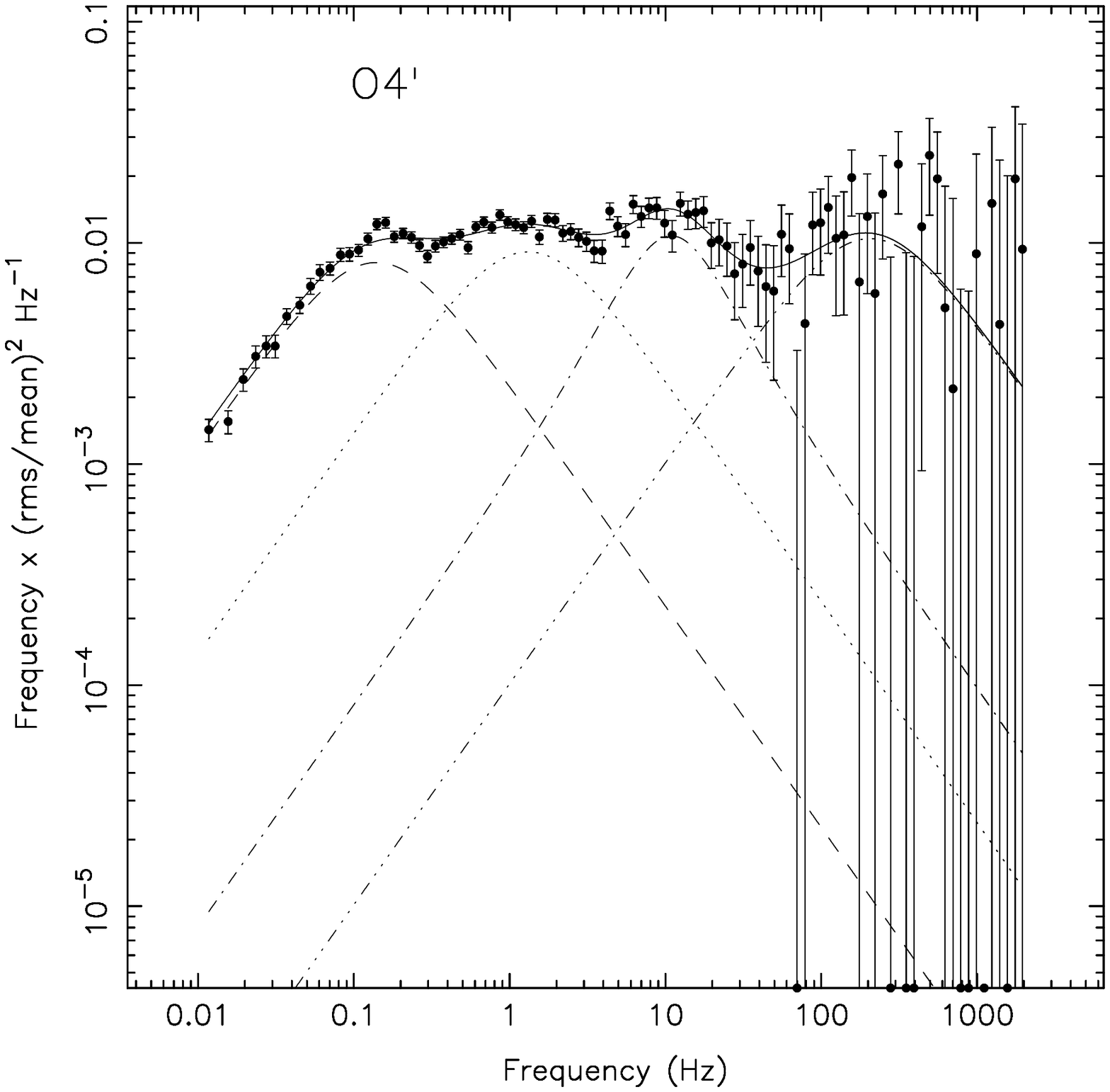}  & 
\plotone{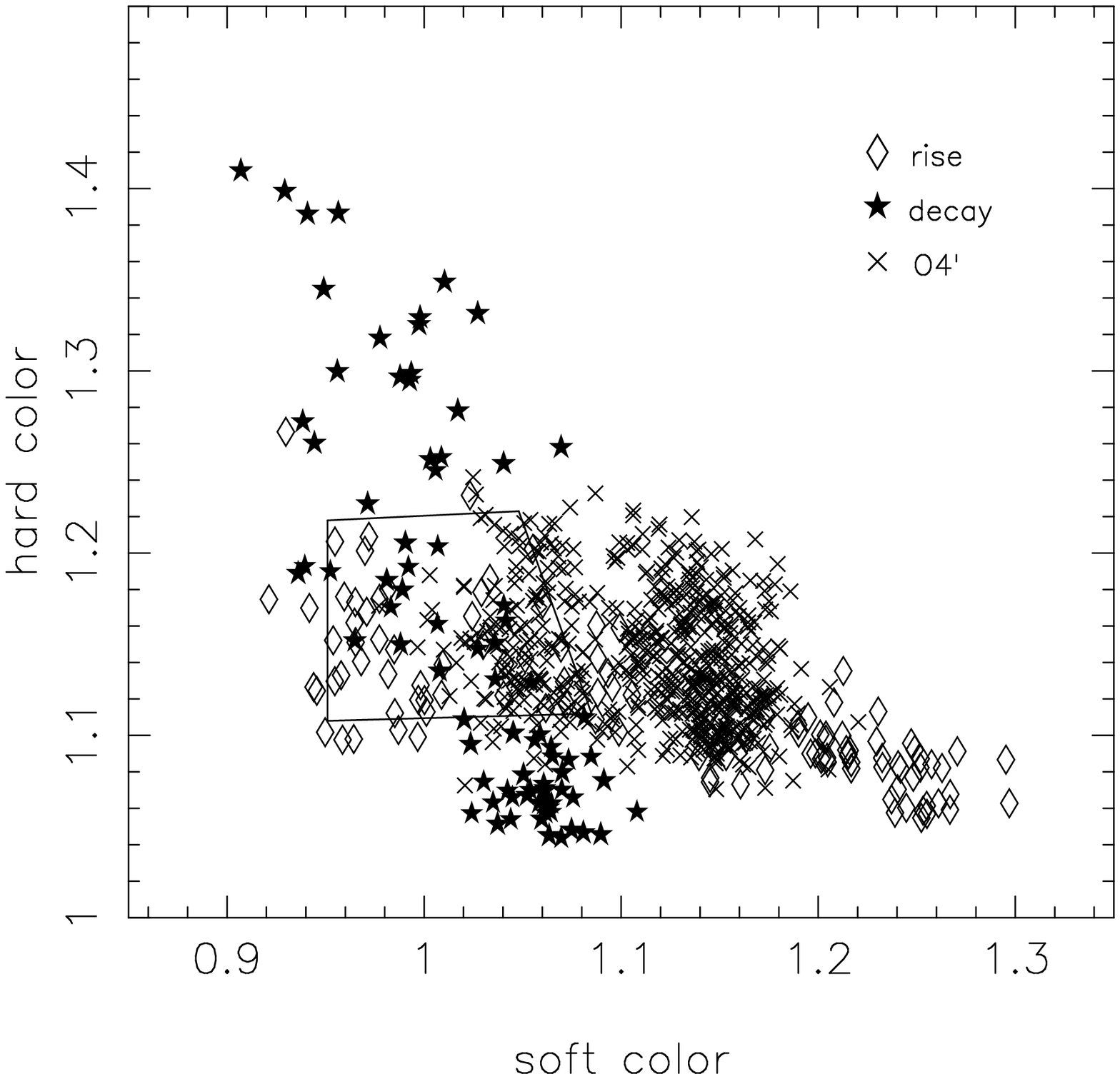} &\\
\end{tabular}
\caption{Comparison of the power spectra for three different parts of the
X-ray outburst. All three power spectra correspond to the same region of 
the CD. An enlarged view of the island state marking the color region is
also plotted.
\label{island}}
\end{figure}

One interesting result from the timing analysis of the extreme island
state is the differences in the aperiodic variability of the source during
the rise and decay of the outbursts. Figure~\ref{island} shows three power
spectra of \aql\ and an enlargement of the CD displaying the extreme
island state. All three power spectra are associated with the same value
of the colors, i.e, they occupy the same position in the CD (marked in
Fig.~\ref{island}). However, they correspond to different parts of the
X-ray outburst: rise (diamonds),  decay (stars) and when the source found
itself in the extended island state of O4$^{\prime}$ (crosses). During the
rise the power is concentrated at low frequencies whereas during the
decay, the source exhibits more power at high frequencies During the
extended island state of outburst O4$^{\prime}$ there is roughly equal
power at all frequencies. The differences in the characteristic
frequencies of the noise components are much less pronounced and
consistent with one onother whenever the same component appears.

\subsubsection{The banana state}

The low-frequency end ($\simless$ 1 Hz) of the banana state power spectra
is dominated by the very-low frequency noise (VLFN). One single power law
is normally used to fit this component. However, the power spectrum of
\aql\ below 1 Hz cannot  be described by one component only.   In order to
fit the VLFN we used two zero-centered Lorentzians that were called low
VLFN ($L_{LVLFN}$) and high VLFN  ($L_{HVLFN}$) (Schnerr et~al. 2003). 
The characteristic frequencies of these components are found to be
independent of the position of the source in the CD: $\nu_{\rm
max}\sim0.01$ Hz and $\sim$0.03 Hz, respectively. However, while the
fractional rms variability of lower frequency VLFN roughly remains
constant at $\sim$3.4\%, the higher frequency VLFN becomes stronger ---
increasing from $\sim$0.9\% in the lower banana to $\sim$2\% in the upper
banana --- as the count rate increases.

Above 1 Hz the banana state power spectra contain one broad Lorentzian,
describing the band-limited noise and one QPO (two in BS1). The
characteristic frequency and width of the broad band-limited noise
component decrease as the count rate increases (i.e. as the source moves
toward the right).  This component has also been seen in 4U 1608--52 (van
Straaten et al. 2003), 4U 0614+09 and 4U 1728--34 (van Straaten et al.
2002). However, the lack of data precludes a more detailed comparison.
Only 4U 1728--34, with just two measurements, seems to have the same
behavior as \aql\, i.e., the frequency of this noise component decreases
along the banana. Following van Straaten et~al. (2003) we will refer to
this component as $L_{b_2}$. Next to, sometimes on top of, the
band-limited noise in the banana state there is a relatively narrow peak,
whose characteristic frequency and width do not correlate with any other
parameter. This narrow QPO has a frequency in between that of the hHz
Lorentzian and $L_h$. van Straaten et~al. (2003) --- see also di Salvo
et~al. (2001) and van Straaten et~al. (2002) --- argued that it is the
band-limited noise component $L_b$ which transforms into a narrow QPO
above certain frequency ($\sim 20$ Hz in 4U 1608--52). The appearance of
$L_{b_2}$ and this transformation of $L_b$ occurs coincidentally. In \aql,
although the narrow version of $L_b$ appears systematically in all
banana-state power spectra it is always below $3\sigma$. 

The detection of a second kHz QPO in \aql\ is only marginal (van der Klis
2000). Based on the values of the fitting parameters and the relation
between the QPO frequency and the X-ray colors, M\'endez, van der Klis \&
Ford (2001) concluded that whenever a kHz QPO is detected in \aql\ it is
always the lower kHz QPO. The fractional  rms and characteristic frequency
of the second QPO in BS1 are consistent with being the lower kHz QPO
(Fig. 1  of M\'endez, van der Klis \& Ford 2001). Also, the characteristic
frequencies of the $L_\ell$ and $L_b$ in BS1 agree with the correlation of
Fig.~\ref{corr} if an extrapolation to higher frequencies of the \aql\
points is done.

\subsubsection{The island state (IS)}

The analysis of the transition state is hampered by poor statistics given
the relatively low number of points. One single Lorentzian with peak at
$\approx 20$ Hz and Q value of 0.15 accounts for the entire 0.01--100 Hz
frequency band. This noise component is identified as $L_b$. A second
narrower Lorentzian with $\nu_{\rm max} \approx 710$ Hz and $Q\approx 3$
fits the noise at higher frequencies. Its fractional rms, $\sim 17$\%, is
too high to agree with the lower kHz QPO. In fact, the value of the
frequency and rms are similar to what it is seen in  4U 1608--52 and 4U
1728--34 for the upper kHz QPO (Fig.~1  of M\'endez, van der Klis \& Ford
2001). Nevertheless, given the relatively noisy spectrum at high
frequencies such identification should be taken with care.

\subsection{The normal/flaring branch-like oscillations}

\begin{figure}
\epsscale{0.75}
\begin{tabular}{ccc}
\plotone{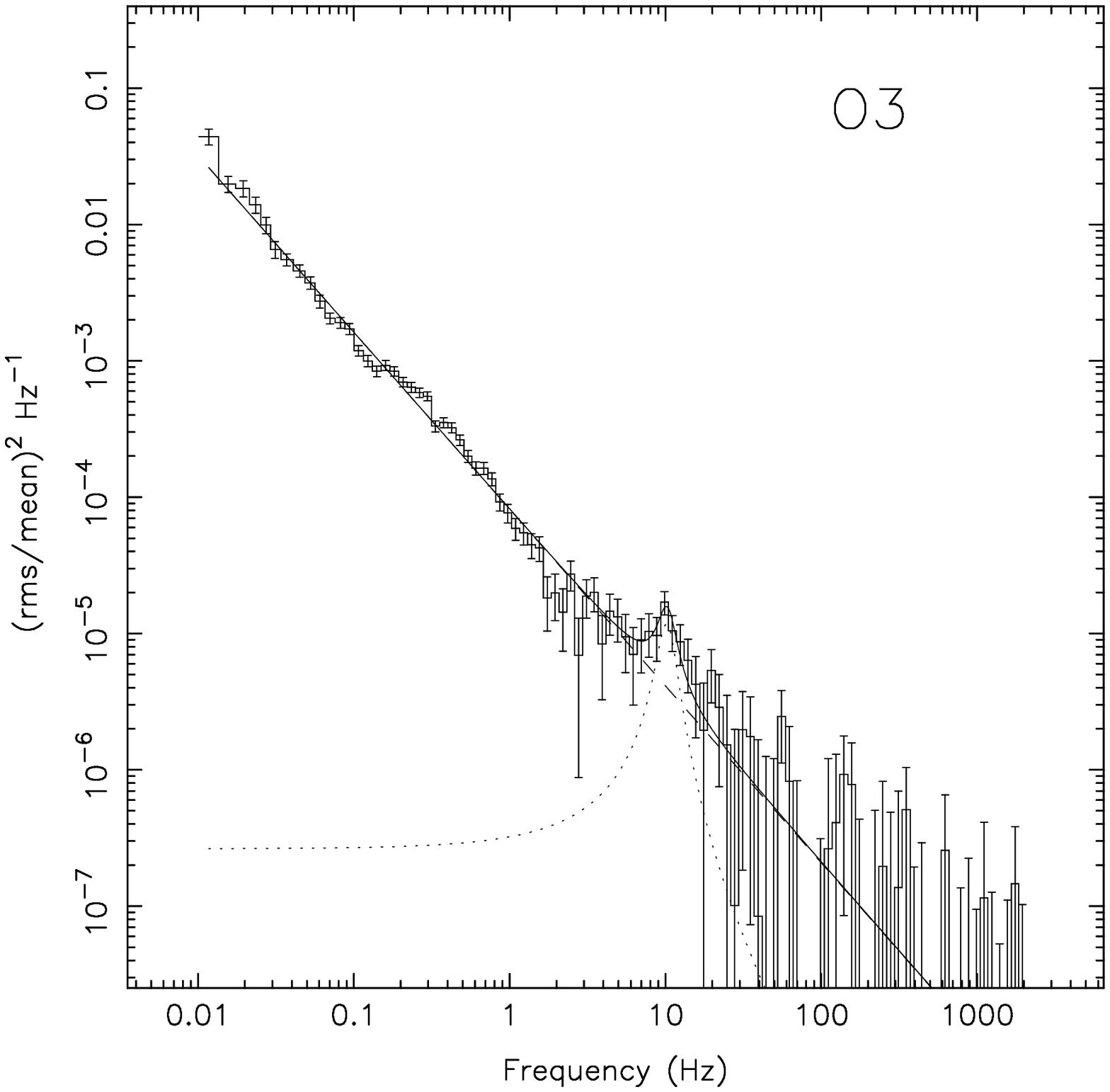}& 
\plotone{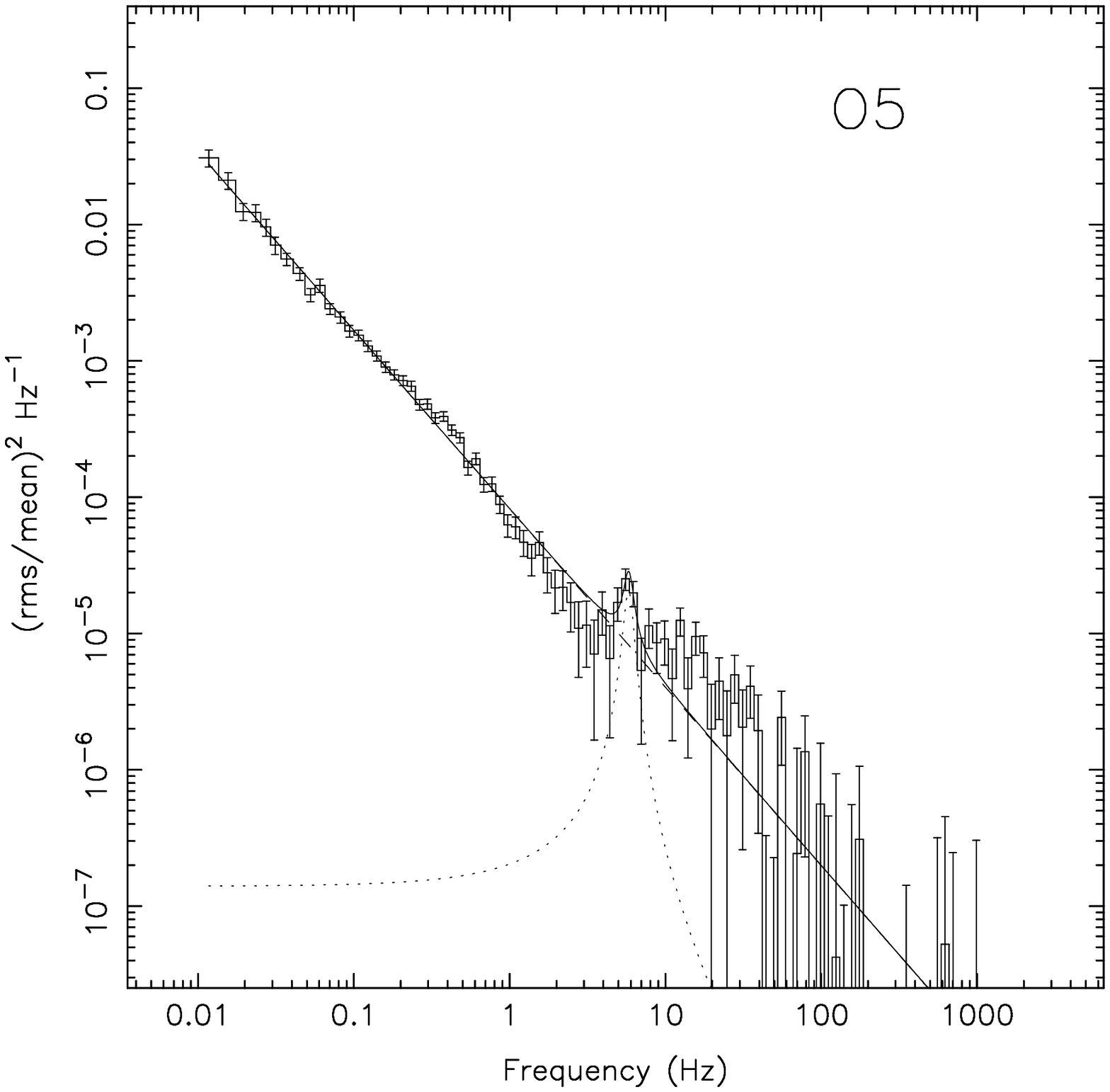}& \\
\end{tabular}
\caption{N/FBO-like feature in the power spectra of \aql\ during the
peak of outbursts 3 and 5.
\label{qpo}}
\end{figure}

A QPO with frequency 7--14 Hz has been found in at least two atoll
sources, XTE J1806--246 (Wijnands \& van der Klis 1999; Revnivtsev,
Borozdin \& Emelyanov 1999) and 4U 1820--30 (Wijnands et al. 1999). These
QPOs are detected at the highest inferred mass accretion rates only, are 
short-lived (a few hundreds of seconds) and are localized in a very narrow
region of the CD, namely the tip of the upper banana. They are reminiscent
of those observed in Z sources in the normal and flaring branches, when
the source is accreting near the Eddington limit. Thus the detection of
similar QPO in atoll sources at significantly lower accretion rates
questions the current models that explain NBO in terms of near-Eddington
accretion (Fortner, Lamb \& Miller 1989). We searched for NBO-like QPO in
\aql\ and tentatively found two of them at $\sim 3\sigma$ significance, at
frequencies 10.3$\pm$0.5 Hz in O3 and 5.8$\pm$0.2 Hz in O5
(Fig.~\ref{qpo}). They have similar rms amplitude ($\approx 0.7$\%) and
Q-values ($\approx 3$). The average values of the colors and intensity 
associated with these QPOs are SC=1.11, HC=0.52, $I_{\rm X}$=5045 count
s$^{-1}$ and SC=1.06, HC=0.51, $I_{\rm X}$=6870 count s$^{-1}$,
respectively. For the sake of comparison with other sources we also give
the value of the X-ray luminosity: $8.5 \times 10^{36}$ erg s$^{-1}$ for
outburst 3 and  $8.1 \times 10^{36}$ erg s$^{-1}$ for outburst 5. As a
confirmation that they occur at the highest count rates, typical
luminosities in the softest parts of the extreme island and banana states
are  $3 \times 10^{34}$ and $2 \times 10^{36}$ erg s$^{-1}$, respectively.
The values of the luminosity are for a distance of 2.5 kpc and the energy
range 2--16 keV.  In the other three outbursts the detection is only
marginal. The low statistical significance of these QPOs ($\simless 2.5$)
can be ascribed to their extremely short life. 

\section{Discussion}
\label{discuss}

We have analysed all the available RXTE data of \aql\ between 1997
February and 2002 May and investigated its spectral and timing properties
in correlation.  This work has been motivated by recent reports that show
that if large data sets are used, the atoll sources displaying the largest
ampitude variations in intensity show three-branch patterns in the
color-color diagram (CD), reminiscent of Z sources (Muno, Remillard \&
Chakrabarty 2002; Gierli\'nski \& Done 2002). Consequently, they conclude
that the distinction between atoll and Z sources might be no longer
justified. The extended island state, island state and banana
state of atoll sources would then correspond to the horizontal branch
(HB), normal branch (NB) and flaring branch (FB), respectively. The
results of these studies were based on the spectral properties (i.e.
color-color diagram) only. In order to be able to distinguish between the
different types of low-mass X-ray binaries the rapid aperiodic variability
associated with the various spectral states needs to be addressed as well
(Hasinger \& van der Klis 1989), and that is what we have attempted in
this paper. We first make a comparison of the timing properties of \aql\
with those of other atoll sources and then discuss whether ot not it can
be considered as a Z source.

\subsection{Comparison with atoll sources}

4U 1608--52 and \aql\ are the atoll sources that display the largest
dynamic range in X-ray intensity ($I_{max}/I_{quiesence} > 1000$). Thus a
comparison between these two sources is of particular interest. 4U
1608--52 has been extensively studied by van Straaten et al. (2003). By
comparing the shape of the power spectra (see Fig.~7 in van Straaten et
al. 2003 and our Fig.~\ref{pds}) and the results of the power spectral
fits (their Table 2 and 3 and our Table~\ref{fitres})  one can  find
distinct similarities between the two sources: the power spectra of
intervals A, B and C in 4U 1608--52 resemble those of the extreme island
state of \aql, while interval D of 4U 1608--52 is reminiscent of the
island state in \aql. Likewise, the shape and number of Lorentzians
appearing in the power spectra of the extreme island state of \aql\ are
similar to those of the three low-luminosity X-ray bursters studied by
Belloni et al. (2002).

It is worth noting the relevant role of the hard color in connection with
the timing properties of the extreme island state.  In \aql\ the
characteristic frequencies of the timing features in the extreme island
state increase from left to right. This is opposite to what it is seen in
4U 1608-52, where frequencies increase from right to left. 
However, in both sources the increase in
frequencies occurs in correlation with an overall decrease in hard color.
We therefore suggest that in the extreme island state it is the hard
color, rather than the intensity, which is the main determining factor of
the timing properties. We note, that black hole candidates are
likewise characterized by power spectral states whose occurrence seems to
primarily depend on spectral hardness and which are relatively insensitive
to intensity, and which are associated with two-dimensional motion through
the color-intensity plane rather than motion along a one-dimensional track
(Homan et al. 2001). It may be that in the island and extreme
island states, the states in which atoll sources are most similar to black
holes (e.g., van der Klis 1994a; Belloni, Psaltis \& van der
Klis 2002), possibly because in these states the inner disk radius is
furthest from the neutron star surface, this 2-D picture better describes
neutron star behavior than the 1-D description appropriate to the banana
state and to Z sources.

Leaving aside the kHz QPOs, the overall shape of the banana-state power
spectra of \aql\ present similarities with those of other atoll sources,
namely, a strong red-noise component, the very-low frequency noise (VLFN)
below $\sim$ 1 Hz and a broad  and a narrow components describing the
band-limited noise (BLN) in the frequency range 1--100 Hz. Compared to  4U
0614+09, 4U 1608-52 and 4U 1728--34, \aql\ exbibits the strongest VLFN but
the weakest BLN. While the VLFN is normally described by either a power
law with index between 1.5-2.0 or a zero centered Lorentzian in the other
atoll sources, it requires the use of two model components in \aql,
similar to the case of GX 13+1 (Schnerr et al. 2003). The broad
band-limited noise component $L_{b_2}$ in \aql\ and 4U 1608-52, or
equivalently the zero centred Lorentzian in 4U 1728--34 and 4U 0614+09
have similar rms values. In contrast, $L_b$, which is clearly detected in
these other atoll sources, is not significant in \aql\ in the banana
state.

\subsection{Is \aql\ a Z source?}

Gierli\'nski \& Done (2002) suggested that the name atoll source is no
longer appropiate  as these type of low-mass X-ray binaries also display a
three-branch pattern in the color-color diagram when large data sets of
observations are used, reminiscent of the Z pattern for which Z sources
are named. This effect is more pronounced in systems exhibiting large flux
variations (Muno et al. 2002), like \aql. The CD of
\aql\ (Fig.~\ref{ccd}) does indeed show a branch (the island state) that
can be interpreted to connect the two main branches like in a Z shape. In
addition, the correlation between the characteristic frequency of the
1--20 Hz broad Lorentzian $L_h$ and the band-limited noise component
(Fig.~\ref{corr}) at low count rates, suggesting a common origin with the
horizontal branch oscillations and LFN in  Z sources and the discovery of
7--14 Hz QPO  at the highest X-ray flux (Fig.~\ref{qpo}), reminiscent of
the Z-source normal/flaring branch oscillations, strengthen the
similarities between \aql\ and Z sources. 

However, there are a number of systematic differences that clearly puts
\aql\ outside the group of Z sources. First, there is the motion in the
CD. In the extreme island state as the count rate increases the soft color
increases and the hard color decreases. Then the source makes a rapid
transition to the  banana state and proceeds to harder colors (right) as
the count rate increases further. However, after the peak of the outburst,
when the intensity decreases the source does not follow exactly the
reverse path. As shown in Fig.~\ref{CDmotion}, \aql\ enters the banana
state (extreme island state) at a higher (lower) value of the soft color
than when it leaves it. The source moves clokcwise following a rectangle
track. It should be pointed out that the entrance into the banana state at
such high soft color is not caused by the time resolution used ($\sim$ 1
day) in Fig.~\ref{CDmotion}. A 256-s bin shows the same effect. However,
we cannot rule out an observational origin since the time gap between the
transition from the extreme island state to the banana state is
$\approx$~5 days for outburst 5 and $\approx$~2.6 for outburst 4
(Table~\ref{timetran}). If this is the case and the source entered the
banana state from the left then the pattern that it would trace out in the
CD would be that of an inverted triangle, much like 4U 1705--44 (Barret \&
Olive 2002). In contrast, the re-entrance into the extreme island state at
a low soft color is probably not cause by the lack of data given the
relatively short gap between the island state and the extreme island state
(just 1.6 days in outburst 5). In either case this behavior is not that of
Z sources. In Z sources, the NB is not a fast transition between the other
two branches, and the source follows the same path whether it is going up
or down along the NB.  In this respect, \aql\ is like the other two
atoll sources that display high-amplitude flux variations. More detailed
looks into the motion of 4U 1608--52 and  4U 1705--44 have revealed
different topologies from the classical Z-shaped track. The motion of 4U
1608--52 resembles the greek letter $\epsilon$ (van Straaten et al. 2003)
and 4U 1705--44 describes an inverted triangle (Barret \& Olive 2002). As
noted above, these various topologies may simply be different aspects of
2-D motion through the color-intensity plane in which it is hard color
that determines the timing properties.

Second, the amplitude of the X-ray luminosity change over the Z track is
also different.  In \aql\ the X-ray luminosity  throughout the CD, i.e.,
from the left of the banana state to the right of the extreme island state
is about three orders of magnitude. In Z sources, luminosity changes are
typically less than a factor of 2 (di Salvo et~al. 2000, 2001, 2002). 

Third, another difference is the velocity and time spent by the source in
each spectral state. Jonker et~al. (2002) calculated that the Z source GX
5--1 spent most of the time in the normal branch. In contrast,  \aql\
spends the smallest percentage of the the total observing time in the
transition state (the analogue to the normal branch would \aql\ be a Z
source).  Wijnands et~al. (1997) found that Cyg X--2 moves through the Z
most slowly on the HB, faster on the NB and fastest on the FB. During a
typical X-ray outburst, i.e., when \aql\ is X-ray active, it spends most
of its time in the banana branch (the analogue to the flaring branch). 

Finally, the properties of the aperiodic variability of \aql\ also differ
from those in Z sources, especially at very low count rates. We can
establish the following differences:

\begin{itemize}

\item [--]The very existence of the extreme island state. Although the
characteristic frequencies in \aql\ increase along the extreme island
state as the count rate increases as Z sources do in the HB, the latter
never  reach the low frequencies that are seen in atoll sources in the
extreme island state, even when at the left end of the HB. Likewise, the
rms amplitude of the noise components in the extreme island state is
significantly higher than in a Z source HB. The total rms amplitude of
those components in \aql\ amounts to $\sim 30$\%, while typically in Z
sources, it does not go above $\sim 10$\%. The extreme island-state power
spectra are typical of atoll sources.

\item [--]The peaked noise components, such as $L_h$ and the upper kHz QPO
are narrow QPOs on the HB in Z sources, with typical values of the Q
parameter above 3 for the upper kHz QPO, whereas they are considerably
broader peaked noise components in the extreme island state in \aql (Q $<$
0.5). 

\item [--]In \aql, the normal/flaring branch-like oscillation does not
occur  during the "normal/flaring branch", i.e. the island or the lower
banana branches but in the upper banana branch, just like in other atoll
sources.

\item [--]The strength of the kHz QPOs decreases as the Z source moves
along the Z track from HB to NB. Typically, the kHz QPOs become
undetectable by the time the source reaches the middle of the NB. In
contrast, \aql\ presents kHz QPOs in the island and banana states.

\item [--]The time scales of the aperiodic variability tend to decrease
(frequencies increase) as the Z source moves along the FB, which is the
opposite behavior to what is is observed in \aql\ in the banana branch.

\end{itemize}

\section{Conclusions}

We have investigated the timing and spectral properties of the soft X-ray
transient \aql\ during five X-ray outbursts covering a period of more than
five years. Three spectral states show up in the color-color diagram of
\aql: a low/hard state, identified with an extreme island state, a
high/soft state corresponding to the classic banana state and an
intermediate state associated with the island state of atoll sources. We
have found that the hard color plays a crucial role in determining the
timing properties of the extreme island state. Although the overall
distribution of these states in the color-color diagram may resemble a Z,
neither the motion in the CD, nor the typical time scales through the
different branches of the CD, nor the range of X-ray luminosities, nor the
timing properties at very low count rates (the existence of the extreme
island state) are compatible with the classical Z-source behavior.

\acknowledgements
PR is a researcher of the programme {\em Ram\'on y Cajal} funded by the
University of Valencia and the Spanish Ministery of Science and
Technology. PR also wants to thank the University of Crete  for providing
in part the resources needed to carry out this work. This research has
made use of data obtained through the High Energy Astrophysics Science
Archive Research Center Online Service, provided by the NASA/Goddard Space
Flight Center.


\clearpage
\begin{deluxetable}{lcccc}
\scriptsize
\tablecolumns{5}
\tablecaption{Duration and intensity
(2--16 keV, 5 PCUs) of the spectral states for
each of the five X-ray outbursts \label{dur}}
\tablewidth{0pt}
\tablehead{}
\startdata
States	&Duration &I$_{\rm min}$ &I$_{\rm max}$ &I$_{\rm mean}$ \\
	&days	&c/s		&c/s		&c/s  \\
\tableline
\multicolumn{5}{c}{Outburst 1} \\
\tableline
EIS	&--/0.14	&--/56		&--/63		&--/60 \\
BS	&15.1		&583		&3057		&2164  \\
IS	&0.10		&200		&220		&209   \\
\tableline
\multicolumn{5}{c}{Outburst 2} \\
\tableline
EIS	&--		&--		&--		&--    \\
BS	&30.96		&1213		&2694		&1778  \\
IS	&--		&--		&--		&--    \\
\tableline
\multicolumn{5}{c}{Outburst 3} \\
\tableline
EIS	&--		&--		&--		&--    \\
BS	&38.95		&1227		&6175		&4656  \\
IS	&--		&--		&--		&--    \\
\tableline
\multicolumn{5}{c}{Outburst 4} \\
\tableline
EIS	&6.02/2.03	&110/70		&1295/156	&650/127 \\
BS	&18.05		&368		&4835		&2330  \\
IS	&--		&--		&--		&--    \\
\tableline
\multicolumn{5}{c}{Outburst 4$^{\prime}$} \\
\tableline
EIS	&63		&77		&530		&340   \\
BS	&--		&--		&--		&--    \\
IS	&--		&--		&--		&--    \\
\multicolumn{5}{c}{Outburst 5} \\
\tableline
EIS	&6.54/1.33	&87/128		&1444/255	&542/203 \\
BS	&41.53		&810		&8555		&4925  \\
IS	&0.01		&458		&463		&460   \\
\tableline
\enddata
\tablenotetext{}{EIS: extreme island, BS: banana, IS: island}
\tablenotetext{}{The left (right) values correspond to the rise (decay) of
the outburst}
\end{deluxetable}
\begin{deluxetable}{lccc}
\scriptsize
\tablecolumns{4}
\tablecaption{Duration in days of the spectral transitions. These values
should be considered as upper limits as they include observational
gaps \label {timetran}}
\tablewidth{0pt}
\tablehead{}
\startdata
Transition	&Outburst 1	&Outburst 4	&Outburst 5 \\
\tableline
EIS$\longrightarrow$BS	&--	&2.63	&4.94   \\
BS$\longrightarrow$EIS	&4.83	&1.06	&2.64   \\
BS$\longrightarrow$IS	&1.89	&--	&0.98   \\
IS$\longrightarrow$EIS	&2.84	&--	&1.64   \\
\tableline
\enddata
\end{deluxetable}

\clearpage

\begin{deluxetable}{lccccccccc}
\scriptsize
\tablecolumns{10}
\tablecaption{Boundaries and properties of the color regions on which the 
timing analysis was performed \label{reg}}
\tablewidth{0pt}
\tablehead{}
\startdata
States		&$(SC,HC)_1$&$(SC,HC)_2$&$(SC,HC)_3$ &$(SC,HC)_4$&Mean	&Mean	&Mean Count 	&Number	of&$\chi^2$(dof)\\
		&	&	&	&			&$SC$	&$HC$	&Rate$^a$ (c/s)	&power spectra	&\\
\tableline
EISrise1	&(0.89,1.08)&(1.12,1.08)&(0.89,1.25)&(1.11,1.26)&1.01	&1.14	&185	&54	&122/92\\
EISrise2	&(1.14,1.06)&(1.21,1.07)&(1.14,1.15)&(1.26,1.14)&1.18	&1.09	&925	&34	&105/84\\
EISrise3	&(1.19,1.04)&(1.34,1.05)&(1.23,1.10)&(1.29,1.11)&1.25	&1.07	&1063	&23	&95/87 \\
EISdecay1	&(1.01,1.02)&(1.11,1.03)&(1.00,1.12)&(1.11,1.23)&1.05	&1.07	&183	&37	&87/92\\
EISdecay2	&(0.92,1.11)&(1.11,1.14)&(0.88,1.43)&(1.06,1.45)&0.98	&1.24	&76	&40	&106/92 \\
EISO4$^\prime_1$&(0.96,1.04)&(1.24,1.06)&(0.96,1.26)&(1.10,1.26)&1.05	&1.15	&188	&152	&106/87\\
EISO4$^\prime_2$&(1.08,1.04)&(1.07,1.06)&(0.96,1.26)&(1.10,1.26)&1.14	&1.14	&356	&367	&114/84\\
IS		&(0.97,0.71)&(1.11,0.89)&(0.82,0.97)&(0.95,1.05)&0.98	&0.90	&255	&28	&99/91\\
BS1		&(0.83,0.47)&(0.93,0.41)&(0.89,0.59)&(0.95,0.57)&0.91	&0.50	&1275	&313	&193/87\\
BS2		&(0.93,0.41)&(0.99,0.40)&(0.95,0.57)&(1.01,0.57)&0.97	&0.49	&2086	&368	&139/89\\
BS3		&(0.99,0.40)&(1.04,0.41)&(1.01,0.57)&(1.02,0.57)&1.02	&0.49	&3507	&368	&100/88\\
BS4		&(1.04,0.41)&(1.10,0.42)&(1.02,0.57)&(1.05,0.61)&1.07	&0.50	&5248	&381	&89/88\\
BS5		&(1.01,0.42)&(1.18,0.50)&(1.05,0.61)&(1.09,0.68)&1.12	&0.51	&5315	&250	&99/88\\
\enddata
\tablenotetext{a}{Background subtracted count rate for the five PCU and 
bandwidth 2--16 keV}
\end{deluxetable}

\clearpage

\begin{deluxetable}{lccccccc}
\scriptsize
\tablecolumns{8}
\tablecaption{Power spectral parameters. 
Errors are 1$\sigma$ ($\Delta\chi^2=1$) \label{fitres}}
\tablewidth{0pt}
\tablehead{}
\startdata
State$^a$ 	&$L_{LVLFN}$&$L_{HVLFN}$ &$L_b$  	&$L_{b_2}$	&$L_h$		&$L_\ell$		&$L_{hHz}$,$L_u$ \\
\tableline
\multicolumn{8}{c}{Characteristic frequency $\nu_{\rm max}$ (Hz)} \\
\tableline
EISrise1		&-	&-	&0.10$\pm$0.01	&-		&0.59$\pm$0.07	&12.1$\pm$1.0	&-\\
EISrise2		&-	&-	&0.44$\pm$0.02	&-		&2.62$\pm$0.06	&21.6$\pm$0.9	&233$\pm$23\\
EISrise3		&-	&-	&0.61$\pm$0.01	&-		&5.4$\pm$0.2	&32$\pm$2	&273$\pm$24 \\
EISdecay1		&-	&-	&2.3$\pm$0.3	&-		&17$\pm$3	&-		&1160$^b$\\
EISdecay2		&-	&-	&0.47$\pm$0.05	&-		&-		&-		&880$\pm$250\\
EISO4$^\prime_1$	&-	&-	&0.130$\pm$0.006&-		&1.11$\pm$0.07	&10.8$\pm$0.8	&180$^b$\\
EISO4$^\prime_2$	&-	&-	&0.284$\pm$0.009&-		&2.17$\pm$0.04	&17.9$\pm$0.6	&223$\pm$17 \\
IS		&-	&-		&20$\pm$2	&-		&-		&-		&700$\pm$40 \\
BS1  &0.0130$\pm$0.0007	&0.36$\pm$0.11	&62$\pm$5	&42$\pm$2 	&-		&861$\pm$6	&- \\
BS2  &0.0144$\pm$0.0009	&0.20$\pm$0.02	&73$\pm$5	&45$\pm$5	&-		&-		&-\\
BS3  &0.0160$\pm$0.0009	&0.25$\pm$0.01	&76$\pm$3	&24$\pm$3	&-		&-		&- \\
BS4  &0.0133$\pm$0.0007	&0.27$\pm$0.01	&73$\pm$4	&16$\pm$1	&-		&-		&-\\
BS5  &0.0103$\pm$0.0010	&0.25$\pm$0.01	&65$\pm$6	&13$\pm$1	&-		&-		&-\\
\tableline
\multicolumn{8}{c}{Integrated fractional rms$^c$} \\
\tableline
EISrise1		&-	&-	&14.8$\pm$0.7	&-		&16.1$\pm$0.6	&19.2$\pm$0.5	&-\\
EISrise2		&-	&-	&13.6$\pm$0.3	&-		&15.8$\pm$0.6	&16.1$\pm$0.6	&11.4$\pm$0.4\\
EISrise3		&-	&-	&11.9$\pm$0.4	&-		&18.2$\pm$0.5	&10.0$\pm$1.0	&12.2$\pm$0.8 \\
EISdecay1		&-	&-	&9.5$\pm$0.6	&-		&13.3$\pm$0.7	&-		&19$\pm$2  \\
EISdecay2		&-	&-	&11.0$\pm$0.4	&-		&-		&-		&38.5$\pm$0.5 \\
EISO4$^\prime_1$	&-	&-	&16.0$\pm$0.4	&-		&15.9$\pm$1.0	&17.9$\pm$1.0	&17.8$\pm$1.9 \\
EISO4$^\prime_2$	&-	&-	&14.9$\pm$0.2	&-		&14.9$\pm$0.5	&17.3$\pm$0.6	&14.0$\pm$0.8\\
IS		&-	&-		&15.7$\pm$0.5	&-		&-		&-		&17$\pm$2\\
BS1	&3.19$\pm$0.06	&0.89$\pm$0.07	&$<$1.3		&5.0$\pm$0.2	&-		&5.4$\pm$0.2	&-\\
BS2	&3.44$\pm$0.05	&1.48$\pm$0.04	&$<$1.2		&3.4$\pm$0.1	&-		&-		&-\\
BS3	&3.49$\pm$0.05	&1.76$\pm$0.03	&$<$1.4		&2.3$\pm$0.1	&-		&-		&-\\
BS4	&3.33$\pm$0.06	&1.97$\pm$0.02	&$<$1.1		&1.79$\pm$0.06	&-		&-		&-\\
BS5	&3.29$\pm$0.11	&2.10$\pm$0.02	&$<$1.0		&1.55$\pm$0.06	&-		&-		&-\\
\tableline
\multicolumn{8}{c}{Q value} \\
\tableline
EISrise1		&-	&-	&0$^b$	&-	&0$^b$		&0$^b$		&-  \\
EISrise2		&-	&-	&0$^b$	&-	&0.42$\pm$0.05	&0.19$\pm$0.07	&0.48$\pm$0.15 \\
EISrise3		&-	&-	&0$^b$	&-	&0.16$\pm$0.05	&0.49$\pm$0.15	&0.43$\pm$0.16 \\
EISdecay1		&-	&-	&0$^b$	&-	&0$^b$		&-		&0$^b$         \\
EISdecay2		&-	&-	&0$^b$	&-	&-		&-		&0$^b$         \\
EISO4$^\prime_1$	&-	&-	&0.14$\pm$0.04	&-&0.22$\pm$0.09&0.15$\pm$0.13	&0$^b$       \\
EISO4$^\prime_2$	&-	&-	&0.03$\pm$0.02	&-&0.33$\pm$0.04&0.16$\pm$0.06	&0.25$\pm$0.13 \\
IS			&-	&-	&0.11$\pm$0.09	&-&-		&-		&1.9$\pm$0.9   \\	
BS1			&0$^b$	&0$^b$	&0$^b$		&5$^b$		&-	&7$\pm$1&-      \\
BS2			&0$^b$	&0$^b$	&0.16$\pm$0.08	&3.5$^b$	&-	&-	&-	\\
BS3			&0$^b$	&0$^b$	&0.23$\pm$0.09	&2.9$\pm$1.2	&-	&-	&-	\\
BS4			&0$^b$	&0$^b$	&0.29$\pm$0.07	&2.4$\pm$1.1	&-	&-	&-	\\
BS5			&0$^b$	&0$^b$	&0.41$\pm$0.08	&2.6$\pm$1.7	&-	&-	&-	\\
\enddata
\tablenotetext{a}{Refers to the regions defined in the color-color diagram
(see Table~\ref{reg})} \\
\tablenotetext{b}{Fixed}
\tablenotetext{c}{Upper limits are at 95\% confidence level}
\end{deluxetable}

\clearpage

\clearpage

\clearpage


\clearpage

\clearpage

\clearpage


\clearpage

\clearpage

\clearpage

\clearpage

\clearpage


\end{document}